\documentclass[smallextended,natbib,runningheads]{svjour3}
\journalname{Annals of the Institute of Statistical Mathematics}
\smartqed  
\usepackage[dvipdfmx]{graphicx}
\usepackage{amssymb}
\DeclareMathAlphabet{\mathpzc}{OT1}{pzc}{m}{it}
\usepackage{colordvi}
\renewcommand{\Blue}{}
\renewcommand{\Red}{}
\newcommand{\RRed}{}
\newcommand{\RRRed}{}
\newcommand{\FRed}{}
\newcommand{\FFRed}{}
\newcommand{\FFFRed}{}
\newcommand{\IBlue}{}
\usepackage{mathptmx}      
\usepackage{amsmath}
\usepackage{amsfonts}
\begin{document}
\title{Multicanonical MCMC for Sampling Rare Events\Red{: \protect \\ An Illustrative Review} 
}



\author{Yukito Iba \and Nen Saito \and Akimasa Kitajima}


\institute{Y. Iba \at
              The Institute of Statistical Mathematics and SOKENDAI, 
10-3 Midori-cho, Tachikawa, Tokyo 190-8562, Japan,
              \email{iba@ism.ac.jp}           
           \and
           N. Saito \at
Research Center for Complex Systems Biology, The University of Tokyo 3-8-1 Komaba, Meguro-ku,Tokyo 153-8902, Japan,
\email{saito@complex.c.u-tokyo.ac.jp}
           \and
            \at
           A. Kitajima \at
              Digital Information Services Division, Digital Information
              Department, National Diet Library,
1-10-1 Nagata-cho, Chiyoda-ku, Tokyo 100-8924, Japan,
 \email{a-kitaji@ndl.go.jp}
}

\date{Received: date / Revised: date}

\maketitle

\begin{abstract}
Multicanonical MCMC (Multicanonical Markov Chain Monte Carlo; Multicanonical Monte Carlo)
is discussed as a method of 
{\it rare event sampling}. Starting from a
review of the generic framework of  importance sampling,
multicanonical MCMC is introduced,  followed by
applications in random matrices, random graphs, and
chaotic dynamical systems. Replica exchange MCMC
(also known as parallel tempering or Metropolis-coupled MCMC) 
is also explained as an alternative to multicanonical MCMC.
In the last section, multicanonical MCMC
is applied to data surrogation;
a successful implementation in surrogating
time series is shown.
\RRed{In the appendices, calculation of averages and normalizing constant 
in an exponential family, phase coexistence, simulated
tempering, parallelization, and multivariate extensions
are discussed.} 
\end{abstract}

\sloppy
\section{Introduction}
\label{intro}
 
Multicanonical MCMC (Multicanonical Markov Chain Monte Carlo; Multicanonical Monte Carlo)~was introduced 
in statistical physics in the early 1990s (\cite{berg1991multicanonical,
berg1992multicanonical,berg1992new}); it can be viewed as a variant 
of umbrella sampling, whose origin can be traced back to the 1970s (\cite{Torrie1974578})~\footnote{
A rarely cited paper, 
\cite{mezei1987adaptive}, 
already proposed an adaptive version of umbrella sampling, which
uses a general ``reaction coordinate'' instead of total energy.
\cite{baumann1987noncanonical} is also referred to as a prototype
of multicanonical MCMC.
}. 
The Wang--Landau algorithm developed in \cite{wang2001efficient,
wang2001determining} provides an effective realization
of a similar idea and many current studies use this
implementation. 
\Red{It, however, relies 
on step-by-step realization of
``multicanonical weight'' defined in Sec.~\ref{sec:mulweight}
in this paper, which is an essential part of the original multicanonical 
\FFFRed{algorithms} by \cite{berg1991multicanonical,berg1992multicanonical} and
\cite{berg1992new}. In this paper, we will use the term
``multicanonical MCMC'' for any method that uses the multicanonical weight.} 

In these studies, multicanonical MCMC is applied to
simultaneous sampling from Gibbs distributions of different temperatures;
in terms of statistics, it corresponds to sampling from an exponential
family. From this viewpoint, a major advantage of multicanonical MCMC  is
fast mixing in multimodal problems. \Red{It often realizes an order of 
magnitude improvement in the speed of convergence over conventional MCMC.}
Some examples in statistical physics are provided by the references in Sec.~\ref{sec:stat_phys}; see also
review articles \cite{berg2000introduction, janke1998multicanonical,
landau2004new, higo2012enhanced, iba2001extended}.

Recent studies, however, provide another look at this algorithm.
Multicanonical MCMC enables an efficient way of sampling
{\it rare events} under a given distribution.
Suppose that rare events of  $x$ in a high-dimensional sample space 
are characterized by the value of statistics $\xi(x)$. Then, in some
examples, rare events even with probabilities $P(\xi_0 \leq\xi(x) ) \approx 10^{-100}$
are sampled within a reasonable computational time~\footnote{
The constant $\xi_0$ controls rareness; see Sec.~\ref{sec:rareimp} for details.}. 
Further, these probabilities are precisely estimated without additional computation.

This novel viewpoint opens the door to a
broad application field of multicanonical MCMC,
while providing a more intuitive and easy understanding of the same algorithm.
Even though some surveys have already introduced 
multicanonical MCMC as a method of rare event sampling
\FRed{(see \cite{driscoll2007searching}; \cite{bononi2009fresh}; \cite{Wolfsheimer2011})}~\footnote{
See also \cite{2012arXiv1212.0534B}; this paper introduced a related
algorithm, split sampling, as a method of rare event sampling.
},  it will be
useful to conduct another survey with a broad perspective and novel applications. 
An aim of this paper is to provide such an introduction, including
recent results by the authors.

Another aim of this paper is 
to apply multicanonical MCMC  to
{\it exact tests} in statistics.
Multicanonical MCMC is useful for sampling
from highly constrained systems, and this will be explained
in this paper in connection with rare event sampling. 
Hence, it can be naturally applied to MCMC exact tests
 (\cite{besag1989generalized, diaconis1998algebraic}), 
where constraints among variables make it difficult to construct Markov chains
for efficient sampling  from null distributions. 
As an example, we will discuss
surrogation of nonlinear time series;
yet the proposed method can be generalized to the other MCMC 
exact tests such as sampling from tables with fixed marginals.
The results discussed in Sec.~\ref{sec:surr} are published 
here for the first time 
in English.

The rest of this paper is organized as follows: In Sec.~2, multicanonical MCMC
is surveyed as a rare event sampling technique. Starting from general issues on
rare event sampling,  the use of an exponential family with replica
exchange MCMC
is discussed as an alternative to multicanonical MCMC. 
Then, the key idea of 
multicanonical MCMC is introduced, and a concise description
of the Wang--Landau algorithm is provided. 
Sec.~3 provides examples of multicanonical 
rare event sampling, focusing on the authors'
recent studies on random matrices, random graphs, and dynamical systems.
Sec.~4 begins with a multicanonical approach
to highly constrained systems. Then, 
exact statistical tests and data surrogation are introduced 
as application fields. A numerical experiment is discussed, where  surrogates of time series 
that maintain the values of  correlation functions are generated. 
\RRed{An appendix deals
with several other issues on multicanonical MCMC, that is,
calculating averages and normalizing constant 
in an exponential family, ``phase coexistence,'' simulated tempering,
parallel computation, and multivariate extensions~\footnote{\RRed{
In this paper, double quotes (`` $\cdots$ '') are used
for marking technical terms in physics, non-technical expressions, and 
terms defined in this paper, 
whereas \FRed{{\it italics} are} utilized 
for emphasizing other terms.}}.}

This paper is mainly intended to describe 
the possibility of multicanonical MCMC in various fields.
Therefore,  we focus on basic concepts and examples, 
omitting details such as
mathematical proofs of convergence and 
practical issues of implementation.
\Red{We assume the readers are familiar with standard algorithms of 
MCMC, but do not have specific knowledge on rare event sampling nor
multicanonical MCMC. Thus, we begin with basics of rare event sampling
and proceed to multicanonical MCMC, skipping details
of the implementation of MCMC. In fact, we can combine almost any kind
of MCMC algorithm to the idea of multicanonical MCMC. It is, however,
essential to pay attention to the behavior of the sample path in the case of 
multimodal distributions, which we will discuss in detail in the paper.
}

\Red{
Readers who are not familiar with MCMC will find necessary
backgrounds, for example, in~\cite{Gilks_Richardson_Spiegelhalter199512, 
RobertCasella200508, Brooks_Gelman_Jones_Meng201105}. 
See also books on MCMC
by physicists, such as~\FFRed{\cite{newman1999monte, FrenkelSmit200111,
Berg200410, LandauBinder201311, BinderHeermann201210}.}}

\section{Multicanonical Sampling of Rare Events}

\subsection{Rare Event Sampling}

We first consider general issues in rare event sampling, namely, 
importance sampling and the use of  exponential families; replica
exchange MCMC is also explained.
For further details on general frameworks and other approaches, see 
\cite{Bucklew200403,RubinsteinKroese200712,Rubino_Tuffin200905}.

\subsubsection{Importance Sampling} \label{sec:rareimp}

Let us assume that the value of a variable 
$X$ is randomly sampled from the probability distribution
$P$; throughout this paper, we assume that $P$ is precisely known. \RRed{
Hereafter, for simplicity, we explain cases where 
variable $X$ takes discrete values; \FRed{however, 
generalization to a continuous $X$
is not difficult.}} 

When we specify target statistics $\xi$,
``rare events'' of $X$ with a rare value $\xi(X)$ of $\xi$ 
are defined as a set \Blue{${\cal A}=\{ x \, | \, \xi_0 \leq \xi(x) \}$},
where the probability $P(\xi_0 \leq \xi(X))$ takes a small value~\footnote{
$P(\xi_0 \geq \xi(X))$ is reduced to the case 
$P(\xi_0 \leq \xi(X))$ by considering $-\xi$, and
hence, it is not discussed separately. The probability
$P(\xi_0-\delta \leq \xi(X) \leq  \xi_0+\delta)$ 
is also considered.
In this case,
we should maintain an adequate value of  $\delta$ and/or
consider the relative probabilities using the same value of 
 $\delta$ for a proper definition of ``rareness.''
}; the constant  $\xi_0$ controls the rareness of the events.

Our problem is to generate samples of $X$ that satisfy
$\xi_0 \leq \xi(X)$ and estimate their probability 
$P(\xi_0 \leq \xi(X))$.
Given current hardware, we can still complete
the task by a direct computation, even when 
the probability $P(\xi_0 \leq \xi(X))$ takes considerably 
smaller values such as $10^{-4}$ or $10^{-6}$.  
However, when the probability of rare events is much smaller, say,
$10^{-12}$ or even $10^{-100}$, it is virtually impossible to deal with
the problem by naive random sampling from the original
distribution $P$. 

A standard solution to this problem is the use of {\it importance sampling}
techniques, that is, we generate samples of $X$ 
from another distribution $Q$, which has a larger probability 
in the set $\mathcal{A}$ . Hereafter, 
we assume that $Q(X=x) \neq 0$ for the value of $x$ satisfying
$P(X=x) \neq 0$. Using samples 
$X^{(i)}, i=1, \ldots, M $ from $Q$, 
the probability under the original distribution $P$ is estimated as
\begin{equation}\label{eq:p1}
P(\xi_0 \leq \xi(X)) \simeq
\frac{1}{M}\sum_{i=1}^M 
\left [ \frac{P\left (X^{(i)} \right )}{Q\left (X^{(i)}  \right )} \,
I(\xi_0 \leq \xi(X^{(i)})) \right ],
\end{equation}
where $I$ is defined by
\begin{equation}\label{eq:defI}
I(\xi_0 \leq \xi(X^{(i)}))= 
\begin{cases} 
1,  & \xi_0 \leq \xi(X^{(i)})   \\  0, &  \xi_0 > \xi(X^{(i)})
\end{cases}.
\end{equation}
By the law of large numbers, 
\eqref{eq:p1} becomes an equality as $M \rightarrow \infty$.
An average of arbitrary statistics $A(X)$ in the set $\mathcal{A}$ with
weights proportional to $P$ is calculated as
\begin{equation}\label{eq:a1}
{\mathbb E}[A(X) \, | \, \xi_0 \leq  \xi(X) ] \simeq
\frac{\displaystyle \frac{1}{M}
\sum_{i=1}^M  \left[ A(X^{(i)}) \, 
\frac{P\left (X^{(i)} \right )}{Q\left (X^{(i)} \right )} 
\, I(\xi_0 \leq \xi(X^{(i)})) \right ]}{P(\xi_0 \leq \xi(X))},
\end{equation}
which also becomes an equality as $M \rightarrow \infty$.

\Red{
A critical issue in importance sampling is the choice
of the distribution $Q$.  Prior to the introduction of MCMC,
there was a severe limitation on the choice of $Q$; this was because
efficient generation of samples is possible only
for a simple $Q$. In contrast, MCMC provides
much freedom in the selection of $Q$. On the other hand,
samples from $Q$ generated by MCMC are usually 
correlated, and such correlation can severely affect the
convergence of the averages. Thus, we should pay attention
to the mixing of MCMC in the choice of $Q$.
}

\subsubsection{Exponential Family and Replica Exchange MCMC}

A strategy~\footnote{A different approach to 
combine importance sampling with MCMC is found in \cite{Botev2013}.},
which we will discuss in this paper,
is to choose $Q$ in the form
$$
Q(x) = \frac{G(\xi(x))\, P(x)}{\sum_x G(\xi(x))\, P(x)},
$$
where $G(\xi)$ is an appropriate univariate function and
$\sum_x$ indicates the sum over
the domain of $x$.
\Red{Multicanonical MCMC belongs to this class.
Here, we will discuss a different choice 
\mbox{$G(\xi)= \exp(\beta\xi)$} as an alternative to 
multicanonical approach; this leads to
\begin{equation}\label{eq:gibbs1}
Q_\beta(x) = \frac{\exp(\beta\xi(x))\, P(x)}{\sum_x 
\exp(\beta\xi(x)) \, P(x)}.
\end{equation}
} 
$Q_\beta$  is interpreted as 
an exponential family with sufficient
statistics $\xi$ and a canonical parameter $\beta$; it is also regarded as 
a Gibbs distribution with energy $-\xi$ and 
inverse temperature $\beta$, when the base measure $P$ is uniform.

Assuming $Q_\beta$ defined by~\eqref{eq:gibbs1}, we can sample 
regions with larger values of $\xi$ by increasing the value of $\beta$.
Thus, in principle, MCMC sampling from $Q_\beta$ with a large value of $\beta$ 
can efficiently generate rare events defined by \Blue{$\xi_0 \leq \xi(x)$}.
When $\beta$ increases, however, \Blue{the set $\cal A$ of $x$ 
defined by $\xi_0 \leq \xi(x)$}
often  almost disconnects, that is, it consists of multiple
``islands''  of $x$ separated by regions with tiny values of $Q_\beta$.
Such a multimodal property of $Q_\beta$ obviously leads to slow convergence
of MCMC.

In many examples, this difficulty is reduced using {\it replica exchange MCMC}, 
which is also known as {\it parallel tempering} or 
{\it Metropolis-coupled MCMC}
 (\cite{2074066, citeulike:606345, HUKUSHIMAKoji1996, iba2001extended}). In this algorithm, 
Markov chains with different values of $\beta$
run in parallel; here, we assume $K$ chains
with $(\beta_1, \beta_2, \ldots, \beta_K)$. 
\FFFRed{Selecting a pair $i$ and $j$ of chains
in a regular interval of steps,} \RRRed{the current values of the 
states $X^*_i$ and $X^*_j$} of chains are
swapped with probability $P_{\rm swap}$ defined as
$$
P_{\rm swap}=
\max \left \{ 1, \frac{Q_{\beta_i}(X^*_j)Q_{\beta_j}(X^*_i)}
{Q_{\beta_i}(X^*_i)Q_{\beta_j}(X^*_j)}     \right \}
=
\max \left \{ 1, \exp((\beta_i-\beta_j)(\xi(X^*_j)-\xi(X^*_i))) \right \}.
$$
Note that the combined
probability \RRRed{$\prod_{k=1}^K Q_{\beta_k}(x_k)$} 
is a stationary distribution
of the Markov chain defined by a combination of 
the original MCMC and the exchange procedure defined above.
This property ensures that replica exchange
MCMC realizes a proper sampling procedure at each value of $\beta$.

\begin{figure}[tb]
\begin{center}
\includegraphics[width=7.8cm]{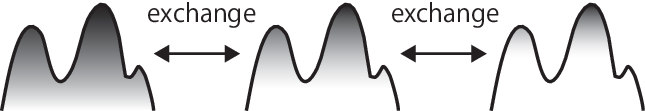}
\caption{
Exchange of states between the distribution
$P_\beta$ with different values of $\beta$. 
The vertical axis corresponds to the value of 
$\xi$, whereas the horizontal axis for each sub-chart 
schematically represents a high-dimensional space of $X$.
Here, the values of 
$\beta$ are assumed to increase from left to right; 
shading represents the changes in 
high-probability regions. 
}\label{fig:replica1}
\end{center}
\end{figure}

Exchange of states between chains is introduced 
for facilitating mixing at large values of $\beta$. 
Owing to these exchanges, states generated 
at smaller values of $\beta$ successively 
``propagate'' to chains with
larger $\beta$ (Fig.~\ref{fig:replica1}).
 This mechanism is similar to that in the
{\it simulated  annealing} algorithm (\cite{kirkpatrick1983optimization})  
for optimization.
An essential difference is that
replica exchange MCMC utilizes a time-homogeneous
Markov chain designed for
sampling from each of the given distributions. 
In contrast, simulated  annealing 
utilizes a time-inhomogeneous chain; at least in principle,
it is not suitable for sampling.

The combination of replica exchange MCMC and 
$Q_\beta$ given by \eqref{eq:gibbs1} provides a powerful 
tool for rare event sampling, which is easy to implement
on parallel hardware. However, the estimation
of the probability of rare events \RRRed{\mbox{$P(\xi_0 \leq \xi(X))$}}
under the original distribution
$P$ requires some additional consideration.
Namely, samples at a single value of $\beta$
are usually not enough for computing relative values of 
\RRRed{\mbox{$P(\xi_0 \leq \xi(X))$}} 
for all values of $\xi_0$. Hence,
the normalizing constant $Z_\beta=\sum_{\, x} \exp(\beta \xi(x))P(x)$ 
should be estimated for combining the results
at different $\beta$.

These difficulties are well treated using  samples at 
multiple values of $\beta$,
which are most naturally
obtained as outputs of replica exchange MCMC. 
\RRed{Here, however, we omit details; essentially,
the same problem in statistical physics is known as the
estimation of ``density of states.'' See, for example, 
an intuitive method used in \cite{hartmann2002sampling} 
and a rather sophisticated approach, the {\it multiple 
histogram method}, explained in \cite{newman1999monte}.}

\subsection{Multicanonical MCMC}

Here, we explain multicanonical MCMC, which is the main
subject of this paper. First, we define a ``multicanonical weight''
and discuss the behavior of MCMC with this weight.
Then, we introduce adaptive MCMC schemes 
for realizing the multicanonical weight. 
\RRed{\FRed{In this section,
we explain the algorithm for cases where both $X$ and $\xi$ take
discrete values.}
A simple way to treat \FRed{a continuous $\xi$}
is introduction of a binning function $\tilde{\xi}$ defined in 
Sec.~\ref{sec:entropic}; for more sophisticated methods,
see references in Sec.~\ref{sec:WL}.}

\subsubsection{Multicanonical Weight} \label{sec:mulweight}

As already explained, when $Q_\beta$ given in 
\eqref{eq:gibbs1} is used,
some additional computation is required for estimating 
the probabilities of rare events. The situation can be
worse in some examples; a region of $\xi$ is virtually not sampled for any choice
of the canonical parameter $\beta$. This may not be typical but possible; 
see Sec.~\ref{sec:co} for further details. 

In contrast, multicanonical MCMC has an advantage in that
it provides probabilities such as  \mbox{$P(\xi_0 \leq \xi(X))$}
directly as outputs of the MCMC simulation
and no additional computation is required. Further, the problem of
the missing region of $\xi$ can be avoided, at least 
in some examples. \RRRed{
In addition to these nice properties, multicanonical MCMC
enables fast convergence in multimodal problems, 
similar to replica exchange MCMC.}

To realize these properties, multicanonical MCMC utilizes
$G(\xi)$ defined in the following way. 
First, we assume that an approximation $\tilde{P}(\xi)$ of $P(\xi)$
is given, in which the marginal probability of $\xi$ is defined as
\RRRed{$P(\xi^\prime)= \sum_{\xi(x)=\xi^\prime}  P(X=x)$, where $\sum_{\xi(x)=\xi^\prime}$
indicates the sum over $x$ that satisfies $\xi(x)=\xi^\prime$.}
Then, $G(\xi)$ is given by
the inverse $1/\tilde{P}(\xi)$ of  $\tilde{P}(\xi)$;  more precisely, we define
\begin{equation} \label{eq:multiw}
G(\xi(x))= \begin{cases} c \, {\tilde{P}(\xi(x))}^{-1}  &
\text{\ if $\xi \in  [\xi_{\min}, \xi_{\max}]$ } \\ \,\,\,\,\,\,\,\, 0 & \text{ else} \end{cases}, 
\end{equation}
where $c$ is an arbitrary constant
and $[\xi_{\min}, \xi_{\max}]$ is an interval $\xi$ of interest.
Note that the values
of $\xi$ that give $P(\xi)=0$ should be excluded 
from the set $[\xi_{\min}, \xi_{\max}]$.
Hereafter, we refer to $G(\xi)$ defined in \eqref{eq:multiw}
as a ``multicanonical weight.'' The corresponding $Q(x)$
is defined as $Q(x)=G(\xi(x))P(x)/C$, where $C=\sum_x G(\xi(x))P(x)$ 
is the normalizing constant; hereafter, the constant $c$ is 
absorbed in $C$ and omitted from the expressions~\footnote{
This $Q(x)$ will also be referred to as 
a ``multicanonical weight'' \FFFRed{on the space of $x$}.
}.

At first sight, the choice of $G(\xi(x))$ shown in \eqref{eq:multiw} 
does not make sense in practice since
the distribution $P(\xi)$ is essentially the one that we want to calculate by
the algorithm. In some cases, we guess a form of $P(\xi)$ and
use it to approximate the multicanonical 
weight (\cite{koerner2006probing, monthus2006probing}),
but this is rather exceptional.
Nevertheless, we leave this question for a while and 
discuss the properties of a multicanonical weight.

\begin{figure}[tb]
\begin{center}
\includegraphics[width=9cm]{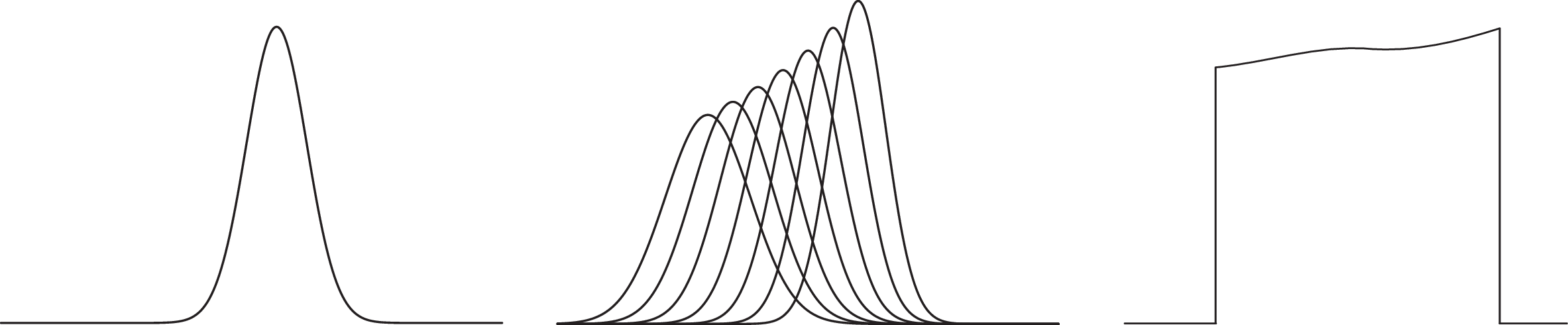}
\caption{
``Flat''  marginal $Q(\xi)$ realized by \eqref{eq:multiw}
is compared to the marginal  
$Q_\beta(\xi)$ of the exponential family
\eqref{eq:gibbs1}.  
Left $Q_\beta(\xi)$ with a fixed value of $\beta$. 
Center a series of  $Q_\beta(\xi)$ with 
$(\beta_1, \beta_2, \ldots, \beta_7)$ are
printed over one another.
Right $Q(\xi)$ 
realized by a multicanonical weight defined in \eqref{eq:multiw};
\Red{a case where $\tilde{P}(\xi) \simeq 
P(\xi)$ is shown, while it becomes completely flat 
when $\tilde{P}(\xi)=P(\xi)$}.
In some cases, behaviors very different from these 
are observed;  see Sec.~\ref{sec:co}.}
\label{fig:multi1}
\end{center}
\end{figure}

Let us tentatively assume an ideal case 
that $\tilde{P}(\xi)$, which appeared in 
the multicanonical weight defined in \eqref{eq:multiw},
is exactly equal to $P(\xi)$.
Then, the marginal distribution $Q(\xi)$ defined by
$Q(x)=G(\xi(x))P(x)/C$ is uniform in the interval 
$[\xi_{\min}, \xi_{\max}]$, excluding the values
of $\xi$ that give $P(\xi)=0$.
This is because the
multicanonical weight is designed for canceling
the factor $P(\xi)$, which is confirmed via direct calculation as
\Red{
\begin{align*}
Q(\xi^\prime)= & 
\frac{1}{C} \sum_x  G(\xi(x)) P(x) I(\xi(x)=\xi^\prime)\\  & 
= \frac{1}{C} G(\xi^\prime) \sum_x P(x) \, I(\xi(x)=\xi^\prime) 
= \frac{1}{C}  \, P(\xi^\prime)^{-1}  \, P(\xi^\prime) =\frac{1}{C} ,
\end{align*}
where $\sum_x$ is the sum over the all possible
values of $x$ and $I(\zeta=\zeta^\prime)$ is defined as
$$
I(\zeta=\zeta^\prime)=\begin{cases} 
1 & \zeta=\zeta^\prime \\ 0 & \zeta \neq \zeta^\prime \end{cases}. 
$$
}
This ``flat'' distribution $Q(\xi)$ of $\xi$ 
realized by a multicanonical weight defined in \eqref{eq:multiw}
is illustrated in the rightmost panel of 
Fig.~\ref{fig:multi1}.  For comparison, $Q(\xi)$ given 
by an exponential family \eqref{eq:gibbs1} is shown in the other two
panels of Fig.~\ref{fig:multi1}.

\subsubsection{MCMC Sampling with a Multicanonical Weight}

So far, we discuss a rather obvious conclusion, but
it is more interesting to consider an MCMC simulation that samples 
the corresponding distribution $Q(x)=G(\xi(x))P(x)/C$. To uniformly 
cover the region $[\xi_{\min}, \xi_{\max}]$, the sample path 
moves randomly in the region. In other words, the multicanonical
weight realizes a random walk on the axis of the target statistics $\xi$; 
this walk has a memory because \RRRed{the value of $X$ does not determined
uniquely by $\xi(X)$}.

\begin{figure}[tb]
\begin{center}
\includegraphics[height=4cm]{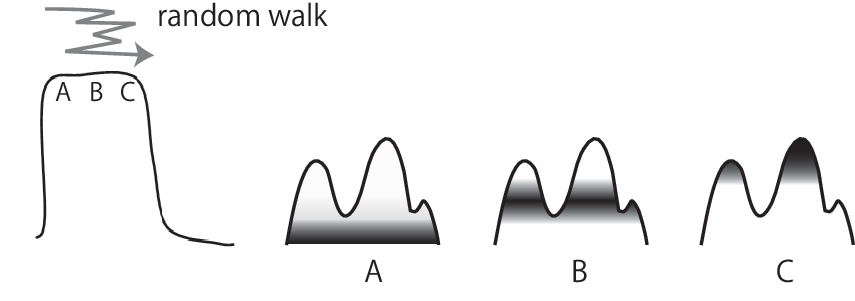}
\caption{\protect\Red{
Random walk  of  $\xi(X)$ realized by a multicanonical weight. 
Left $P(\xi)$ and a sample path of the random walk
projected on the axis of the target statistics $\xi$.
Note that the statistics $\xi$ is a function of state $X$, and 
MCMC updates of $X$
naturally cause such a walk of $\xi$; no
separate procedure is required for changing $\xi$.
Right \FFFRed{A, B, and C show the distributions of $(X,\xi(X))$},
each of which corresponds to the regions A, B, and C in the left panel.
The vertical axis corresponds to the value of 
$\xi$, whereas the horizontal axis for each sub-chart
schematically represents the high-dimensional space \FFFRed{where $X$
takes its value}.
}
}\label{fig:peak2}
\end{center}
\end{figure}

This behavior enables us to obtain the desired 
properties using a single chain, 
as shown in Fig.~\ref{fig:peak2}. 
First, efficient sampling of a tail region with a large value of $\xi$ 
is possible if we choose a sufficiently large $\xi_{\max}$. On the other hand,
fast mixing of MCMC is attained \Red{if  
we choose $\xi_{\min}$ such that \Blue{the set 
defined by $\xi_{\min} \leq \xi(x)$} is tightly connected and 
a sample path can easily move around in it
}~\footnote{
Such a region corresponds to a  ``high-temperature'' region in statistical physics, whereas the 
tail region with  rare events corresponds to a ``low-temperature'' region.}.
Therefore, MCMC sampling with a multicanonical weight 
shares an ``annealing'' property with replica exchange MCMC.

Finally, we confirm how probabilities of rare
events are computed under the original distribution $P$. 
We assume that $X^{(i)}, i=1, \ldots, M $ are samples 
from $Q$ defined by $G$ of the equation \eqref{eq:multiw}. Then,
the following expression is derived from 
\eqref{eq:p1}:
$$
P(\xi_0 \leq \xi(X) \leq  \xi_{\max}) \, \simeq \,
 C \times \frac{1}{M}\sum_{i=1}^M \left [
\tilde{P}( \xi (X^{(i)} )) \,
I(\xi_0 \leq \xi (X^{(i)})) \right ],
$$ 
where $I$ is defined by \eqref{eq:defI}.
Because the values of $\xi$ are limited in
$\xi_{\min}  \leq \xi \leq  \xi_{\max}$
by our definition of the multicanonical weight,
$$ 
P(\xi_{\min}  \leq \xi(X) \leq  \xi_{\max})
\simeq  C \times\frac{1}{M}\sum_{i=1}^M 
\tilde{P}(\xi (X^{(i)})),
$$ 
also holds.
Hence, we arrive at
\begin{equation} \label{eq:multi_est_p}
\frac{P(\xi_0 \leq \xi(X) \leq  \xi_{\max})}{P(\xi_{\min}  \leq \xi(X) \leq  \xi_{\max})} \simeq 
\frac{\sum_{i=1}^M 
\left [ \tilde{P}( \xi (X^{(i)})) \,
I(\xi_0 \leq \xi(X^{(i)})) \right ]}
{\sum_{i=1}^M \tilde{P}( \xi(X^{(i)}))}.
\end{equation}
The value of the denominator $P(\xi_{\min}  \leq \xi(X) \leq  \xi_{\max})$  
becomes almost unity when the interval $[\xi_{\min} , \xi_{\max}]$
contains most of the probability mass; otherwise, in some cases, we are mainly
interested in relative probabilities.   
The expectation of arbitrary statistics $A$
in the tail region $\xi_0 \leq  \xi(x) \leq  \xi_{\max}$ is  
also derived from  \eqref{eq:a1} in a similar manner as
\begin{equation} \label{eq:mul_a1}
{\mathbb E}[A(X) \, | \, \xi_0 \leq  \xi(X) \leq  \xi_{\max}] \simeq 
\frac{\sum_{i=1}^M \left [ A(X^{(i)}) \,
 \tilde{P}(\xi (X^{(i)})) \,
I(\xi_0 \leq \xi (X^{(i)})) \right ]}
{\sum_{i=1}^M 
\left [ \tilde{P}( \xi (X^{(i)} ) ) \,
I(\xi_0 \leq \xi(X^{(i)})) \right ]}.
\end{equation}

\subsubsection{Entropic Sampling} \label{sec:entropic}

Now, we return to the following problem. How to estimate the multicanonical
weight $G(\xi)$ in \eqref{eq:multiw} without prior knowledge?
The key idea is to use adaptive Monte Carlo; ``preliminary runs'' 
of MCMC are repeated to tune the weight $G(\xi)$ until 
the marginal distribution $Q(\xi)$ becomes almost flat in
the interval $[\xi_{\min}, \xi_{\max}]$. After tuning the weight, 
a ``production run'' is performed, where $G(\xi)$ is fixed; this run
realizes MCMC sampling with a multicanonical weight. 
Note that virtually any type of MCMC can be used for sampling
in both of these stages.

An important point is that  $G(\xi)$ is a univariate
function of a scalar variable $\xi$, while \FFFRed{$Q(x)=G(\xi(x))P(x)/C$ 
is defined on a high-dimensional space of $x$}; 
thus, tuning $G(\xi)$ is much easier than performing a direct 
adaptation of \Blue{$Q(x)$} itself. 

To illustrate the principle,  we describe a simple method, sometimes 
known as {\it entropic sampling} (\cite{lee1993new}).
First, we consider the histogram $H$ of the values of $\xi$. 
It is convenient to
introduce a discretized or binned version $\tilde{\xi}(x)$ of $\xi(x)$, which
takes an integer value 
$\tilde{\xi} \in \{1, 2, \ldots, N_b\}$~\footnote{\Red{
Giving a partition $\mathcal{F}_i,i\in \{1, 2, \ldots, N_b\}$
of the interval $[\xi_{\min}, \xi_{\max}]$, \RRRed{it is defined by
\FRed{$\tilde{\xi}(x)=j  \Leftrightarrow \xi(x) \in  \mathcal{F}_j$.}}
If $\xi_{\max}<\xi$
or $\xi < \xi_{\min}$, it is often convenient 
to define \FFFRed{$\tilde{\xi}=N_b$ or
$\tilde{\xi}=1$}, respectively. Another way is to reject the value of
$x$ that satisfies $\xi_{\max}<\xi(x)$
or $\xi (x)< \xi_{\min}$ within the 
Metropolis--Hastings algorithm \FFFRed{(See also a remark in \cite{2003PhRvE}.)}.}}.

Then, the histogram of 
the values of $\tilde{\xi}$ in the $k$th iteration of the preliminary runs
is represented by $\{H^{(k)}(\tilde{\xi})\}, \tilde{\xi}=1,\ldots, N_b$.
We define $\bar{H}$ as expected counts in each bin of a flat
histogram, which is the target of our 
adaptation~\footnote{The constant factor $\bar{H}$ 
is not essential in the following argument when we consider relative weights,
but we retain it because it clarifies the meaning of formulae.}.  
Further, the weight in the $k$th iteration is represented by
$\{G^{(k)}(\tilde{\xi})\}, \tilde{\xi}=1,\ldots, N_b$.
Now that the adaptation in the $k$th step is expressed as
a recursion 
\begin{equation} \label{eq:ent}
G^{(k+1)}(\tilde{\xi})=G^{(k)}(\tilde{\xi})\times 
\frac{\bar{H}\,+\,\epsilon}{H^{(k)}(\tilde{\xi})+\epsilon}.
\end{equation}
Here, a constant $\epsilon$ is required for eliminating the divergence
at  $H^{(k)}=0$, which is set to a small value, say, unity.
The idea behind this recursion is simple---increase
the weight if the counts are smaller than $\bar{H}$ and
decrease the weight if the counts are larger than $\bar{H}$.

The tuning stage of the algorithm is formally described as follows.
Here, we use $LG(\tilde{\xi})=\log G(\tilde{\xi})$ instead
of $G(\tilde{\xi})$.
\begin{enumerate}
\item Initialize $LG$ and set parameters.
\begin{itemize}
\item Set $LG(i)=0$ for $i=1, \ldots, N_b$.
\item \FFRed{Set the maximum number of iterations $K_{\max}$.} 
\item \FFRed{Set the number of MCMC steps $M_{\max}$ within each iteration.}
\item \FFRed{Set the number of MCMC steps $M_s$ between histogram updates.}
\item Set a regularization parameter $\epsilon$ (e.g., $\epsilon=1$).
\item Set $\bar{H}=(M_{\max}/M_s)/N_b$.
\item \FFRed{Set the counter of iterations $K$ to $0$.}
\end{itemize}
\item Initialize $H$ and $X$.
\begin{itemize}
\item Set  $H(i)=0$  for $i=1, \ldots, N_b$. \label{alg:histclear1}
\item Initialize the state $X$.
\item \FFRed{Set the counter of MCMC trials $M$ to $0$.}
\end{itemize}
\item Run MCMC.
\begin{itemize}
\item Run $M_s$
steps of MCMC with 
\RRRed{the weight $P(x)\exp[LG(\tilde{\xi}(x))]$.} \label{alg:MCMC1}
\end{itemize}
\item Update the histogram $H$.
\begin{itemize}
\item \RRRed{$H(\tilde{\xi}(x^*))=H(\tilde{\xi}(x^*))+1$, where 
$X=x^*$ is the current state. $\spadesuit$ } 
\item $M=M+M_s$.
\item If  $M < M_{\max}$, go to Step~\ref{alg:MCMC1}. 
\end{itemize}

\item Check whether $H$ is ``sufficiently flat.''
\begin{itemize}
\item If so, {\bf end}.
\item If not and  $K < K_{\max}$, modify $LG$.
\begin{itemize}
\item \RRRed{$LG(i)=LG(i) + 
\log [ (\epsilon+\bar{H})/(\epsilon+H(\Red{i)})]
$ for $i=1, \ldots, N_b$.  $\clubsuit$}
\item $K=K+1$. 
\item Go to Step~\ref{alg:histclear1}.
\end{itemize}
\item If not and $K_{\max} \leq K$, the algorithm {\bf  fails}.
\end{itemize}
\end{enumerate}

Note that the update formula~\eqref{eq:ent} is included as a step 
marked with $\clubsuit$, while the histogram is incremented
in the step marked with $\spadesuit$. 

After completing the above procedure,
the production run is performed.
If the above algorithm fails to converge, we
can increase the numbers $M_{\max}$ and/or $K_{\max}$.
Another choice is to reduce our requirement and decrease the value
of $\xi_0$, which determines the rareness of the obtained events.   

The construction of the histogram can be replaced by other density estimation
techniques. In the original studies (\cite{berg1991multicanonical, berg1992multicanonical, berg1992new}), 
$\log G(\xi)$ is represented by 
a piecewise linear curve, instead
of a piecewise constant curve used in entropic sampling; parametric 
curve fitting is also utilized. 
\RRed{Another useful method is kernel density estimation,
which is particularly convenient in continuous and/or multivariate $\xi$
cases; it is also used with the Wang--Landau algorithm 
explained \FRed{later}, as seen in \cite{zhou2006wang}}.  
\Red{Finally,
we mention methods based on the {\it broad histogram equation}.
In these methods,
the number of transitions between states are used for optimizing the
weight, instead of the number of visits to a state. Such an idea has
a somewhat different origin (\cite{OLIVEIRA1998}), but it can be interpreted 
as a way to realize a multicanonical weight;
see~\cite{wang2002transition}.}

\subsubsection{Wang--Landau algorithm} \label{sec:WL}

Entropic sampling is already
sufficient for realizing a multicanonical weight in many problems.
In current studies, however, the Wang--Landau algorithm 
(\cite{wang2001efficient,wang2001determining})
is often utilized, which provides a more efficient strategy to construct
a multicanonical weight. 

An essential feature of the Wang--Landau algorithm  is 
the use of a time-inhomogeneous chain in the preliminary runs;
that is to say, we change the weights after each trial of MCMC moves
instead of changing them only at the end of each iteration consisting of
a fixed number of MCMC steps.  
This may lead to  an ``incorrect'' MCMC sampling in the preliminary runs, 
but it causes no problem if we fix the weights in the final
production run, where we compute the required 
probabilities and expectations. 

In the actual implementation, \RRRed{whenever a state $x$ with
$\tilde{\xi}^*=\tilde{\xi}(x)$ appears, we multiply the
value of weight $G(\tilde{\xi}^*)$ by a constant factor $0<C<1$~\footnote
{\RRed{Do not confuse this 
$C$ with the normalization constant $C$ in the
previous sections.}};} 
it reduces the weights of the already visited values of 
$\tilde{\xi}$, whereas it effectively
increases the relative weights of the other values of $\tilde{\xi}$.
In parallel,  we construct the histogram $H$ of $\tilde{\xi}$ that appeared
in MCMC sampling.  After some steps of MCMC, 
we reach a ``sufficiently flat'' histogram~\footnote{\RRed{ 
Usually, in the Wang--Landau algorithm, 
this criterion for flatness should be severer than the requirement 
\FRed{on the flatness
of the histogram} expected in the final production run.}}; then, a step of iterative
tuning of the weights is completed.

When we rerun MCMC where the weight is fixed to 
the values obtained by this procedure, 
the run usually does not provide 
a sufficiently flat histogram of $\tilde{\xi}$.
Then, an iterative method is introduced, that is, 
we increase the value of the constant $C$
and repeat the procedure in the preceding paragraph.
A heuristics proposed in the original papers (\cite{wang2001efficient,
wang2001determining}) 
is to change $C$ to $\sqrt{C}$. 
After each iteration step, the histogram $H$ is cleared, whereas
the values of  $G$ are retained.

Again, we stress that any type of MCMC 
can be used for sampling at each of these stages;  
we use the familiar Metropolis--Hasting
algorithms in the examples considered in this paper. 
As shown in later sections, however, the choice of moves
in the Metropolis--Hasting algorithms significantly affects the
efficiency of the entire algorithm. \\

The tuning of the weight by the Wang--Landau algorithm
is summarized as shown below.  Again, we use
 $LG(\tilde{\xi})=\log G(\tilde{\xi})$ in place of $G(\tilde{\xi})$;
further, we define $LC=-\log C$ (i.e., with a minus sign).

\begin{enumerate}
\item Initialize $LG$ and $LC$; set other parameters.
\begin{itemize}
\item Set $LG(i)=0$ for $i=1, \ldots, N_b$.
\item Set $LC>0$ (e.g., $LC=-\log(1/e)=1$).
\item \FFRed{Set the maximum number of
iterations $K_{\max}$ (e.g., $K_{\max}\!\!=15$ or $18$). }
\item \FFRed{Set the maximum 
number of MCMC steps $M_{\max}$ within each iteration.}
\item \FFRed{Set the counter of iterations $K$ to $0$.}
\end{itemize}
\item Initialize $H$ and $X$.
\begin{itemize}
\item If $K > K_{\max}$, {\bf end}.
\item Set  $H(i)=0$ for $i=1, \ldots, N_b$. \label{alg:histclear}
\item Initialize the state $X$.
\item \FFRed{Set the counter of MCMC trials $M$ to $0$.}
\end{itemize}
\item Run MCMC.
\begin{itemize}
\item Run a 
step of MCMC with the weight $P(x)\exp(LG(\tilde{\xi}(x)))$. \label{alg:MCMC}
\end{itemize}
\item Modify $LG$ and update the histogram $H$.
\begin{itemize}
\item  \RRRed{$LG(\tilde{\xi}(x^*))=LG(\tilde{\xi}(x^*))-LC$,
where $X=x^*$ is the current state.  $\clubsuit$}
\item \RRRed{$H(\tilde{\xi}(x^*))=H(\tilde{\xi}(x^*))+1$,
where $X=x^*$ is the current state. $\spadesuit$}
\end{itemize}
\item Check whether $H$ is ``sufficiently flat.''~\footnote{
\RRed{In actual implementation, this 
step need not to be performed after each step of MCMC;
it can be done, for example, each time after trying to update
all random variables.}}   
\begin{itemize}
\item If so, $LC=LC/2$, $K=K+1$ and go to Step~\ref{alg:histclear}
\item If not and  $M < M_{\max}$, $M=M+1$ and go to Step~\ref{alg:MCMC}.
\item If not and $M_{\max} \leq M$, the algorithm {\bf fails}.
\end{itemize}

\end{enumerate}

Note that update $\clubsuit$ of 
the weight $G$ 
and increment $\spadesuit$ of the histogram $H$
are done simultaneously, in contrast to
entropic sampling. 

\Red{The criterion for a
``sufficiently flat'' histogram used
in Secs.~\ref{sec:largest} and \ref{sec:graph}
is that counts in every bin of the histogram
are larger than 92\% of the value expected
in a perfectly flat histogram. In the cases of Sec.~\ref{sec:graph}, 
we exclude ``permanently'' zero count bins from the criterion, 
where true probability seems zero; it is usually 
difficult to know {\it a priori} and 
some trial and error is required.}

After completing the above  procedure,
the production run is performed.
If this algorithm does not converge or the production run
using the obtained weights does not give a flat histogram of $\xi$, 
what can we do? One possibility is to change the criterion that 
the histogram $H$ is \FRed{``sufficiently flat;''} when we make it more
strict and increase the value of $M_{\max}$, convergence may 
be attained with increasing computational time.  Increasing
the value of $K_{\max}$ may not be effective when we use the original
$\sqrt{C}$ rule for modifying $C$ because the value of $C$ becomes
nearly unity for large $K$. Another possibility is to relax 
our requirement on the rareness
and decrease the value of $\xi_0$.

The algorithm presented here still
contains a number of ad hoc procedures and 
should be manually adapted to a specific problem.
It, however, provides
solutions to problems otherwise difficult to treat.
On the other hand, many modifications of the algorithm
are proposed. Examples of treating continuous variables are
seen in \cite{yan2002density, shell2002generalization, 
liang2005generalized, zhou2006wang, atchade2010wang}.
The following authors have criticized the 
$\sqrt{C}$ rule and have proposed modified
algorithms:  \cite{belardinelli2007wang, 
PhysRevE.75.046701, liang2007stochastic, PhysRevE.78.046705, 
atchade2010wang}.
The convergence of the algorithms is analyzed in
\cite{lee2006convergence, belardinelli2007wang},
while rigorous mathematical proofs are discussed in
\cite{atchade2010wang, 2011arXiv1110.4025J,2012arXiv1207.6880F}.
\cite{2011arXiv1109.3829B} proposed an
automatic procedure including the adaptation of step and bin size. 

\subsubsection{Variance of Estimators}
\Red{
Finally, we will briefly discuss the variance of the estimators.
Here, we restrict ourselves to the final production run with a fixed 
weight. An experimental study
on convergence of estimates is shown in Sec.~\ref{sec:largest}}.

\Red{  
At first, we assume that all samples are independent, although it is not
true for samples generated by MCMC. Then, variances 
of the numerator and denominator of the right-hand side of
\eqref{eq:multi_est_p} are
estimated as
\begin{align} \label{eq:multi_var1}
\sigma_m^2=
\frac{1}{M} \left \{ 
\sum_{\xi=\xi_0}^{\xi_{\max}} \tilde{P}(\xi) \, P(\xi)
- \left [ P(\xi_0 \leq \xi) \right ]^2 
\right \}
\\ \label{eq:multi_var2}
\sigma_{m0}^2=
\frac{1}{M} \left \{
\sum_{\xi=\xi_{\min}}^{\xi_{\max}} \!\!\! \tilde{P}(\xi) \,
P(\xi)
-\left [ P(\xi_{\min} \leq \xi 
\leq \xi_{\max}) \right ]^2  
\right \}.
\end{align}
From \eqref{eq:multi_var1} and \eqref{eq:multi_var2},
the relative variance of the right-hand side of \eqref{eq:multi_est_p}
is estimated as~\footnote{
Here, we apply the delta method using an approximation
\FRed{$\left. \frac{a+\delta a}{b+\delta b} \middle /  
\frac{a}{b} \right. \simeq 1+\frac{\delta a}{a}-\frac{\delta b}{b}$; correlationbetween the denominator and the numerator is ignored
}.
}
\begin{equation} \label{eq:relative_v}
\frac{\sigma_{m}^2}{[P(\xi_{0} \leq \xi)]^2}
+
\frac{\sigma_{m0}^2}{[P(\xi_{\min} \leq \xi \leq \xi_{\max})]^2}.
\end{equation}
}

\Red{
In the case of MCMC, sample correlation becomes important
and we should modify these formulae.
Let us define integrated auto correlation of statistics $B(X)$ as 
\FRed{
$$
\mathcal{T}_B=\frac{1}{\sigma^2_B} \sum_{\tau=1}^{\infty} \left \{ 
\mathbb{E_{\mathrm{path}}} [\, B(X^{(0)}) \, B(X^{(\tau)}) \,]
- \mathbb{E} [\, B(X) \,]^2
\right \},
$$
}
where the expectation 
$\mathbb{E_{\mathrm{path}}}$ 
indicates an 
average over sample paths $X^{(0)}, X^{(1)}, \ldots$ 
generated by MCMC, and $\sigma^2_B$ is the variance
of independent samples from the same distribution. 
Then, the effective number of samples
changes from $M$ to $M/\mathcal{T}_B$, when we calculate
the average of $B$. If we define $\mathcal{T}_m$ and $\mathcal{T}_{m0}$ 
as $\mathcal{T}_B$ with $B(X)=\tilde{P}(\xi(X))  
\,I(\xi_0 \leq \xi(X))$ and $\tilde{P}(\xi(X))$, 
respectively, \eqref{eq:relative_v} is substituted for
\begin{equation} \label{eq:relative_v_corr}
\frac{(1+2 \mathcal{T}_m) \sigma_{m}^2}{[P(\xi_{0} \leq \xi)]^2}
+
\frac{(1+2 \mathcal{T}_{m0}) \sigma_{m0}^2}{[P(\xi_{\min} \leq \xi \leq \xi_{\max})]^2}.
\end{equation}
}

Unfortunately, it is rarely possible to estimate 
$\mathcal{T}_m$ and  $\mathcal{T}_{m0}$ {\it a priori}.
Expression (12), 
however, suggests that 
variances $\sigma_{m}^2$, $\sigma_{m0}^2$ of 
independent samples and integrated auto correlations 
$\mathcal{T}_m$, $\mathcal{T}_{m0}$ \Red{are both important}
in rare event sampling using MCMC. 
\Red{The} multicanonical weight provides
a practical method for balancing them.

If correlation among samples is ignored,
a reasonable choice of $Q$ for sampling from
${\cal A}=\{ x \, | \, \xi_0 \leq \xi(x) \}$ is
$Q_{*}(x)=\tilde{C}P(x)I(\xi_0 \leq \xi(x))$,
which corresponds to the generation of samples
using MCMC from the tail $\xi_0 \leq \xi(x)$ 
of the distribution $P$. It is, however, not useful 
in most practical problems, because it is difficult to 
design a Markov chain that efficiently samples
from $Q_{*}(x)$~\footnote{
In fact, even when conventional MCMC can produce
samples of rare events from $Q_{*}(x)$, 
calculation of the normalizing constant $\tilde{C}$ and 
the probability of rare events are not 
straightforward. An advantage of
multicanonical MCMC is that it provides
a way to calculate the 
probability using \eqref{eq:multi_est_p}.
}.

\section{Examples of Rare Event Sampling by Multicanonical MCMC}

Here, we discuss two applications of
multicanonical MCMC, rare event sampling
in random matrices and chaotic dynamical systems.
Other applications in physics, engineering, and statistics are
briefly surveyed.

\subsection{Rare Events in Random Matrices} \label{sec:random_mat}

A pioneering study on rare events in random matrices
with multicanonical MCMC
is \cite{driscoll2007searching}, which
computes large deviation in {\it growth ratio}, 
a quantity relevant to the numerical difficulty in treating matrices. 
The results in this subsection are discussed in detail in
\cite{saito2010multicanonical} and \cite{SaitoI11}. 
\cite{Kumar2013} also applied the Wang-Landau algorithm
to random matrices using coulomb gas formulation.

\subsubsection{Largest Eigenvalue}\label{sec:largest}

Distributions of the largest eigenvalue $\lambda^{\max}$ 
of random matrices are 
of considerable interest in statistics,
ecology, cosmology, physics, and engineering. 
Small deviations have been studied in this problem, and have yielded
the celebrated  $N^{1/6}$ law by 
\cite{tracy1994level, tracy1996orthogonal}.
Here, we are interested in the numerical estimation of large deviations;
the present analytical approach to large deviations is
limited to specific types of distributions (\cite{dean2008extreme,majumdar2009large}).
Specifically, the probability 
\mbox{$P(\lambda^{\max} < 0)$} that all eigenvalues are negative
is important in many examples, because it is often related to 
the stability of the corresponding systems (\cite{may,aazami2006cosmology}). 

In \cite{saito2010multicanonical}, multicanonical MCMC
is applied to this problem.  Rare events whose probability 
\mbox{$P(\lambda^{\max} < 0)$} is as small as $10^{-200}$ are
successfully sampled for matrices of size 
$N \leq 30$ (or $40$)~\footnote{ 
The most time-consuming part of the proposed algorithm 
is the diagonalization procedure required
for each step of MCMC; the Householder method is used here. 
It can be improved by the use of 
a more efficient method for calculating the eigenvalue $\lambda^{\max}$.}.

\begin{figure}[tb]
\begin{center}
\includegraphics[width=7.2cm]{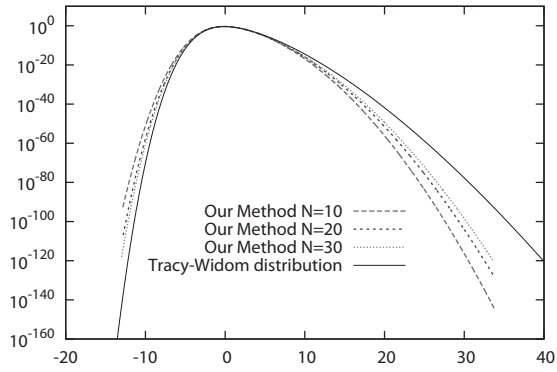}
\caption{The probability
\mbox{$P( \lambda^{\max}_0 < \, \lambda^{\max} \! \leq 
\lambda^{\max}_0 \! +
\delta )$} estimated by the proposed method is plotted
for $N=10, 20$, and $30$, where $\delta$ is a small binsize.
\FRed{The Tracy--Widom distribution for 
{\it small deviation} asymptotics
is shown by the solid curve; systematic deviations from the
obtained result for {\it large
deviation} are observed as expected.}
The horizontal axis corresponds to the scaled variable
$(\,\lambda^{\max}_0-\mathbb{E}(\lambda^{\max}_0)\,)N^{1/6}$.
Gaussian orthogonal ensemble (GOE) is assumed.
[from N.~Saito, Y.~Iba, and K.~Hukushima, Multicanonical sampling of rare events in random matrices, Physical Review E 82, 031142 (2010), 
\copyright~2010 American Physical Society]
}
\label{fig:matrix1}
\end{center}
\end{figure}

\begin{figure}[tb]
\begin{center}

\begin{minipage}{5.3cm}
\includegraphics[width=5.3cm]{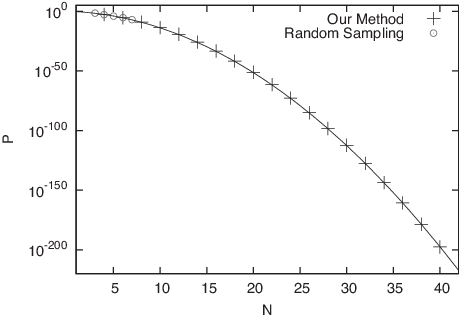}
\end{minipage}
\hspace{0.3cm}
\begin{minipage}{5.3cm}
\includegraphics[width=5.3cm]{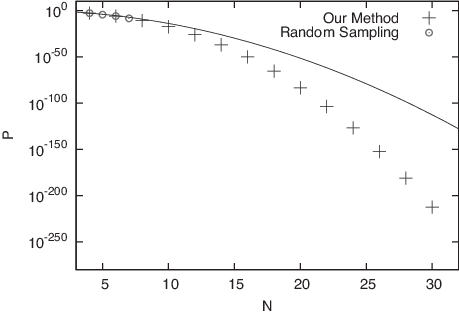}
\end{minipage}

\caption{Probabilities \mbox{$P(\lambda^{\max} <
0)$} obtained by the proposed method and
simple random sampling method are shown against $N$; 
the latter is available only for small $N$.
Left GOE; curve indicates a quadratic fit to the results with Coulomb gas
representation (\cite{dean2008extreme}).
Right an ensemble
of matrices whose components are uniformly distributed;
curve indicates the probability for GOE with
the same variance.
[from N.~Saito, Y.~Iba, and K.~Hukushima, Multicanonical sampling of rare events in random matrices, Physical Review E 82, 031142 (2010),
\copyright~2010 American Physical Society]
}\label{fig:matrix2}
\end{center}
\end{figure}

Examples of the results in \cite{saito2010multicanonical} 
are shown in Figs.~\ref{fig:matrix1} and 
\ref{fig:matrix2}. In Fig.~\ref{fig:matrix1}, the probability 
\mbox{$P( \lambda^{\max}_0 < \, \lambda^{\max} \! \leq 
\lambda^{\max}_0 \! +
\delta )$} is plotted against the values of  
$\lambda^{\max}_0$ with a small binsize $\delta$ \Red{for the case
of Gaussian orthogonal ensemble (GOE).
\FRed{
GOE is defined as an ensemble of random real
symmetric matrices such that entries are independent Gaussian
variables; hereafter, the variances of
the diagonal and off-diagonal components are
1 and 0.5, respectively, while means are all zero.
}
}

Fig.~\ref{fig:matrix2} shows the probability 
\mbox{$P(\lambda^{\max} < 
0)$} that all eigenvalues are negative. 
\Red{The results for GOE and an ensemble
of \FRed{real symmetric matrices}} 
whose components are uniformly distributed
(hereafter ``uniform'') are shown~\footnote{
\FRed{The support of the uniform distributions is chosen
as having the same variance as GOE.}}. For small
$N$s, the results from the proposed method 
reproduce those by simple random sampling~\footnote{ 
\Red{Hereafter, ``simple random sampling''
refers to the method wherein a large number of matrices are independently 
generated from the ensemble and the empirical proportion
is used as an estimator.}}. On the other hand, for a large $N$ for which
simple random sampling hardly suffices,
the obtained results match theoretical results in the case of GOE.
The typical number of steps 
in preparing the multicanonical weight is 
$2 \sim 5 \times 10^9$, and the length of the final productive
run ranges from $1 \times 10^9$ (GOE $N=20$) to $2.5 \times 10^9$
(GOE $N=40$, uniform $N=30$)~\footnote{\RRed{Hereafter, the length of MCMC runs
is measured by the number of Metropolis--Hastings trials; we do {\it not} 
use physicists' \FRed{``Monte Carlo steps (MCS),''} which is defined as the 
number of trials divided by the number of random variables.}}.

\begin{figure}[tb]
\begin{minipage}{6cm}
\centering
\includegraphics[width=6cm]{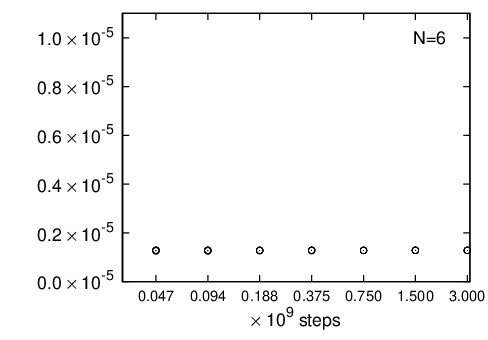}
\end{minipage}
\begin{minipage}{6cm}
\centering
\includegraphics[width=6cm]{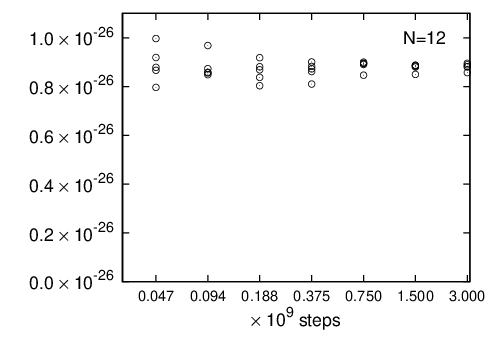}
\end{minipage}
\begin{minipage}{6cm}
\centering
\includegraphics[width=6cm]{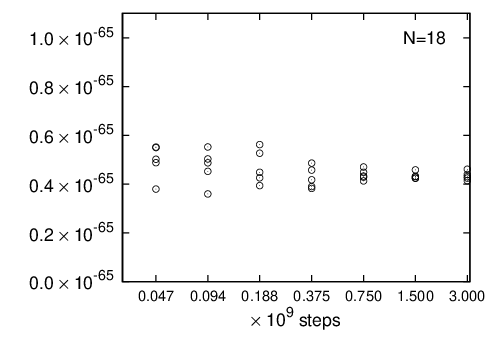}
\end{minipage}
\begin{minipage}{6cm}
\centering
\includegraphics[width=6cm]{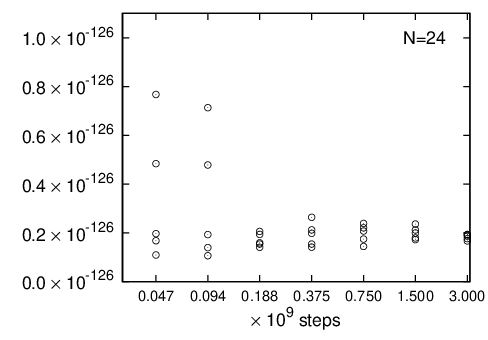}
\end{minipage}
\caption{\RRed{Examples of the convergence of estimated probabilities.
Upper Left~\FRed{$N=6$}, Upper Right~$N=12$, 
Lower Left~$N=18$, and Lower Right~$N=24$. 
The vertical axis corresponds to probabilities \mbox{$P(\lambda^{\max} <
0)$}, while the horizontal axis (log-scale) 
corresponds to the steps of the algorithm. Results of five 
\FRed{production} runs
with the same weight and different random numbers 
are shown for each $N$; in the $N=6$ case, symbols
are almost overlapped one another.
An ensemble of matrices whose components are 
uniformly distributed is assumed.
}}
\label{fig:conv_onlylastran}
\end{figure}

\Red{
Examples of convergence of estimates are shown 
in Fig.~\ref{fig:conv_onlylastran}. \FRed{For an ensemble
of matrices
whose components are uniformly distributed,
multicanonical weights for $N=6,12,18$ and $24$  
are calculated 
by the Wang--Landau algorithm using at most
$5.0\times10^9$ steps.} Then, five independent
production runs are performed 
\FRed{for each $N$ using the same weight obtained 
by this procedure.}
The results for an increasing length of
the production run are shown
in the figure. Noting that the vertical axis of 
Fig.~\ref{fig:matrix2} is log-scale, the variance of
the estimates attained in Fig.~\ref{fig:conv_onlylastran}
is reasonably small and is enough for providing an 
accurate test for asymptotics.} 

The proposed method is quite general and can be applied
to random matrices whose components are sampled from an arbitrary
distribution, or even random sparse matrices, to which
no analytical solution is available. These 
results are discussed in detail in \cite{saito2010multicanonical}, 
along with the detailed specifics of the proposed algorithm.

An important lesson from this example is that we should be
careful while choosing the moves in the Metropolis--Hasting algorithm.
If we generate candidates using 
conditional distributions of the original
distribution, such as \RRed{the Gaussian distribution} 
for each component in GOE, the algorithm fails in some cases.
This occurs because such a method cannot generate
candidates with very large deviations in a
component. This difficulty is avoided by the use of 
a random walk Metropolis algorithm with an adequate
step size; \Red{in the example of GOE, we use \FRed{Gaussian
distributions as proposal distributions in the Metropolis algorithm
(variances are unity for diagonals and 0.5 for 
non diagonals, respectively)}; see \cite{saito2010multicanonical}.}

\subsubsection{Random Graphs} \label{sec:graph}

The search for rare events in random graphs is also an interesting subject.
An undirected graph is represented by the corresponding adjacency
matrix, whose components take values in the set $\{0,1\}$.
For a k-regular graph, the maximum eigenvalue
takes a fixed value equal to $k$, and hence it is not interesting. 
\RRed{On the other hand,
the spectral gap $\lambda^{\rm gap}$,
given as the difference \FRed{between the maximum and
the second-largest eigenvalue} in the case of regular graphs, is related
to many important properties
of the corresponding graph.} Specifically, graphs with larger
values of the spectral gap are called 
{\it Ramanujan graphs} or {\it expanders}; Ramanujan graphs have interesting 
properties for communications and dynamics on networks
(see references in \cite{donetti2006optimal, SaitoI11}).

In earlier studies,
\cite{donetti2005entangled,donetti2006optimal}
optimized the spectral gaps of graphs by simulated annealing; in their
algorithm, a pair of edges of the graph is modified in each 
Metropolis--Hasting step. Using this method,
they showed that expanders with interesting structures automatically
appear. 

\begin{figure}[bt]
\begin{center}
\includegraphics[width=6.5cm]{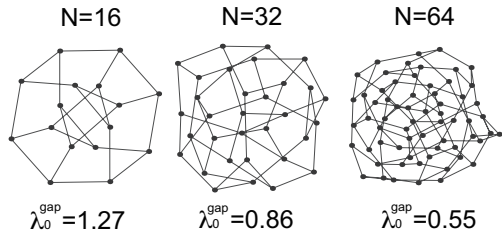}
\caption{Examples of 
3-regular graphs with a large spectral gap found in the
simulation. [from N.~Saito and Y.~Iba, 
Probability of graphs with large spectral gap by multicanonical Monte Carlo, Computer Physics Communications
  182 223-225 (2011), \copyright~2011 Elsevier]}\label{fig:graph1}
\end{center}
\end{figure}

\begin{figure}[bt]
\begin{center}
\includegraphics[width=8.5cm]{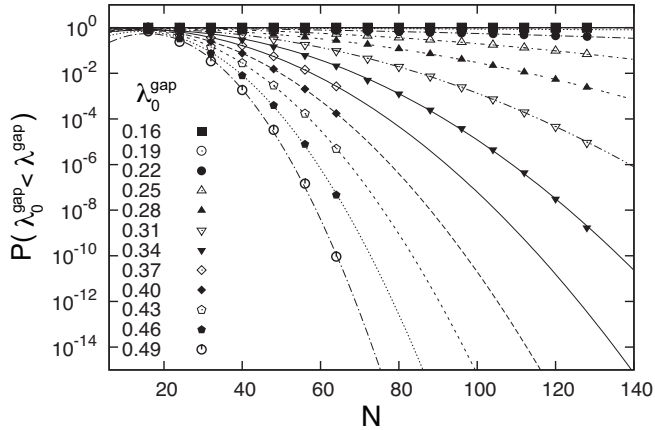}
\caption{
Estimated  \mbox{$P(\lambda^{\rm gap}_0 < 
\lambda^{\rm gap})$}  are shown as functions of $N$. Each
curve corresponds to different values of $\lambda^{\rm gap}_0$. 
Data are well fitted by
quadratic functions when $\lambda^{\rm gap}_0 \gtrsim  0.18$.
[from N.~Saito and Y.~Iba, 
Probability of graphs with large spectral gap by multicanonical Monte Carlo, Computer Physics Communications
  182 223-225 (2011), \copyright~2011 Elsevier]
}\label{fig:graph2}
\end{center}
\end{figure}

\cite{SaitoI11}
applied multicanonical MCMC to this problem; 
they defined the Metropolis-Hasting update as in 
\cite{donetti2005entangled,donetti2006optimal} and used 
the Wang--Landau algorithm for realizing multicanonical weights.
Examples of the obtained graphs are shown in Fig.~\ref{fig:graph1}, while
Fig.~\ref{fig:graph2} gives probability \mbox{$P(\lambda^{\rm gap}_0 < 
\lambda^{\rm gap})$} as a function of $\lambda^{\rm gap}_0$ 
and the size $N$ of matrices. \Red{The typical number of Metropolis 
steps used in preparing multicanonical weights
is $0.5 \sim
1.0 \times 10^9$, while the length of the final production run
is $0.25 \sim 0.5 \times 10^9$.
See \cite{SaitoI11} for further details.}

\subsection{Rare Events in Dynamical Systems}

Rare events in deterministic dynamical systems are important both
in theory and application (\cite{Ott02,BeckSchloegl199501}). 
An example is a quantitative study on tiny tori
embedded in a ``chaotic sea'' of Hamiltonian dynamical systems,
which is a familiar subject in this field. 
Numerical effort required for uncovering these tiny structures 
dramatically increases with the dimension of the system. 
Therefore, it is natural to introduce MCMC and other stochastic sampling 
methods to this field. Studies on MCMC
search for unstable structures in dynamical systems are found in
\cite{Sasa06, yanagita2009exploration, geiger2010identifying}, 
and references therein~\footnote{Sequential Monte Carlo-like algorithms 
are also used; see \cite{tailleur2007probing, 2013arXiv1302.6254L}, and references therein.}.

\cite{KitajimaI11} applied multicanonical MCMC to the study of 
dynamical systems.
In the proposed algorithm,
a  measure of the chaoticity of a trajectory is defined as 
a function of the initial condition~\footnote{
Here, the chaoticity is defined as the number
of iteration required for the divergence of perturbed trajectories; 
the algorithm according to this definition is stable on finite precision machines.}, which corresponds to statistics representing rareness.
Then, the Metropolis--Hastings update is defined as
follows: (1)~perturb the initial condition, 
(2)~simulate a fragment of
trajectory from the new initial condition, and (3) calculate the chaoticity 
of the trajectory and
reject/accept the new initial condition using the current weight. 
Then, the entire algorithm is defined as 
multicanonical MCMC with the Wang--Landau 
algorithm for tuning the weight. 

Again, the choice of moves in the Metropolis--Hastings algorithm
is important; here, we sample a perturbation to the initial
conditions from a mixture of uniform densities 
 with different order of widths. This idea, taken from
\cite{Sweet01}, seems essential for sampling from fractal-like densities;
see \cite{KitajimaI11} for details. 

\begin{figure}[tb]
\begin{center}
\includegraphics[width=9cm]{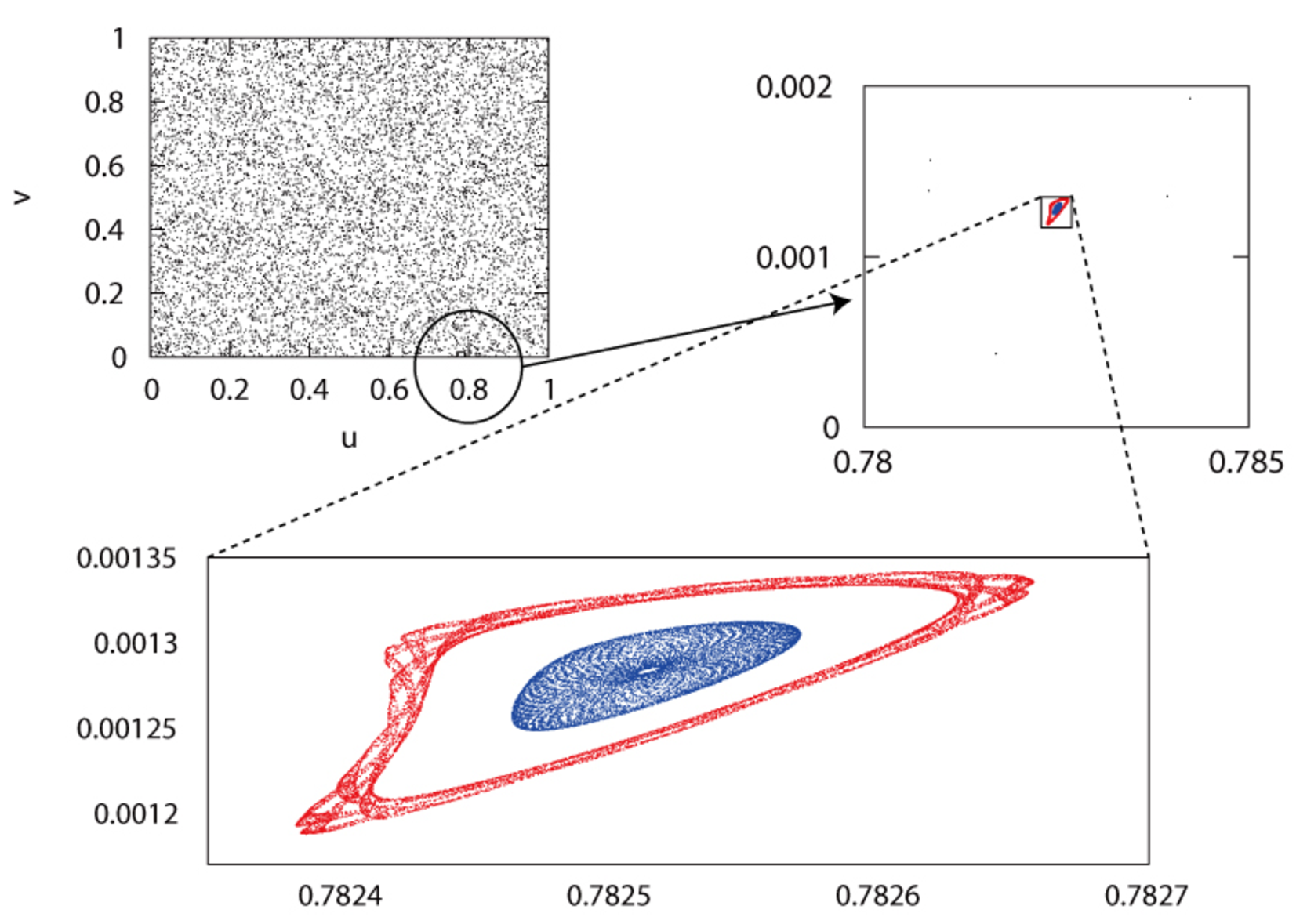}
\caption{Pair of tiny tori in the chaotic sea found by the proposed
method. Projections on the $(u_n, v_n)$-plane are shown. Enlargement
of a tiny area in the small circle in the left panel is given in the right
panel; further enlargement is given in the lower panel. $K = 7.8$ and
$b = 0.001$.
[from A.~Kitajima and Y.~Iba, 
Multicanonical sampling of rare trajectories in chaotic dynamical systems, Computer Physics Communications
  182 251-253 (2011), \copyright~2011 Elsevier]
}
\label{fig:chaos1}
\end{center}
\end{figure}

In \cite{KitajimaI11},
sampling of tiny tori in the chaotic
sea of a four-dimensional map
\begin{eqnarray*}
  u_{n+1} &= u_n-\frac{K}{2\pi}\sin(2\pi v_n)+\frac{b}{2\pi}\sin(2\pi(v_n+y_n)) \\ 
  v_{n+1} &= v_n+u_{n+1} \\
  x_{n+1} &= x_n-\frac{K}{2\pi}\sin(2\pi y_n)+\frac{b}{2\pi}\sin(2\pi(v_n+y_n))\\
  y_{n+1} &= y_n+x_{n+1}
\end{eqnarray*}
is studied,
where $K$ and $b$ are constants that characterize the map.
An example of tiny tori found by the proposed method is shown in
Fig.~\ref{fig:chaos1}. 
\Red{In this case, the total number of initial conditions tested 
in the proposed algorithm is about $4\times10^9$, while
the probability to find an initial configuration
leading to a trajectory with the same degree of 
chaoticity  is as small as $10^{-12}$, assuming
 random sampling \FFFRed{from the Lebesgue measure.}}
In addition, the relative volume of
initial conditions that lead to trajectories of the given order of ``chaoticity''
are successfully estimated by the algorithm; that is, the proposed
method is not only useful for the search but also provides quantitative
information on rare events in dynamical systems, see Fig.~2 of 
\cite{KitajimaI11}.

\subsection{Other Applications}

The rest of this section briefly describes
other fields of applications of multicanonical  MCMC.

\subsubsection{Statistical Physics} \label{sec:stat_phys}

Multicanonical MCMC was originally developed for sampling
from Gibbs distributions in statistical physics.
Hence, a number of studies in this field have successfully applied it to
problems where simple MCMC is virtually 
disabled by slow mixing.  Some typical examples are 
studies on the Potts  and 
other classical spin 
models (\cite{berg1992multicanonical, wang2001determining, 
zhou2006wang}),
spin glass models (\cite{berg1992new, wang2001determining}), and
liquid models (\cite{yan2002density, shell2002generalization, 
calvo2002sampling}).
Multicanonical MCMC is also used for the study of biomolecules 
(see \cite{Mitsutake01, higo2012enhanced} for full-atom protein models
and \cite{chikenji1999multi, wust2012optimized} for lattice protein models).
Attempts to combine the idea of multicanonical weight with chain growth algorithms are found in \cite{bachmann2003multicanonical,prellberg2004flat}.
More examples are found in the review articles mentioned 
in Sec.\ref{intro}.

On the other hand, the use of multicanonical MCMC
for other types of rare event sampling in physics is a recent 
challenge~\footnote{In terms of
physics, it corresponds to the sampling of the ``quenched disorder,''
whereas conventional applications in physics deal with
sampling from the Gibbs distribution of thermal disorder.}.
\cite{hartmann2002sampling} introduced the idea of rare event sampling
by MCMC to the physics community. \cite{koerner2006probing} and 
\cite{monthus2006probing} applied MCMC to the sampling
of disorder configurations that gives
large deviations in ground state energies; in
these studies, modifications of the Gumbel
distribution are used for approximating multicanonical weights. 
Subsequently, \cite{hukushima2008monte} and  \cite{matsuda2008distribution}
applied  the Wang--Landau algorithm to the study 
of Griffiths singularities in random magnets, which is
known to be sensitive to rare configurations of impurities;
\cite{PhysRevE.82.021902} discussed RNA secondary structures.
In these studies, any prior knowledge on the 
functional form of  $\tilde{P}(\xi)$ is assumed. 

\subsubsection{Optical Telecommunication and Related fields}

Multicanonical MCMC
is intensively used for rare event sampling 
in optical telecommunication and related fields.
After a pioneering work by \cite{Yevick2002}, a number of applications
appeared; see, for example, \cite{holzloohner2003use},
and a recent review, \cite{bononi2009fresh}. 
The sampling of rare noises that cause failures of error correction is
discussed in~\cite{holzlohner2005evaluation} and
\cite{yukito2008testing}, which can be useful for
predicting the performance 
of error-correcting codes.

\subsubsection{Statistics}

\Red{Algorithms based on the multicanonical weight,
specifically, the Wang--Landau algorithm and its generalizations, 
increasingly attract the attention of statisticians.}
\cite{liang2005generalized}
introduced the Wang--Landau algorithm to statistics.
\cite{atchade2010wang} and \cite{chopin2012free} 
developed closely related
algorithms and tested them in examples of 
Bayesian inference and model selection. \cite{2011arXiv1109.3829B}
and \cite{Kastner2013244}
also discussed applications in Bayesian statistics;
\cite{Kwon2008} treated a target tracking problem.
\cite{yu2011efficient} (also \cite{LiangLiuCarroll201009})
discussed hypothesis testing 
using stochastic approximation Monte Carlo.
\cite{Wolfsheimer2011} extended
the study of \cite{hartmann2002sampling} 
and applied rare event sampling using
the Wang-Landau method to the computation of p-values for local sequence
alignment problems.
5. Add the following reference to the reference lis
In the following section, we will discuss exact
tests and data surrogation as an application field
of multicanonical MCMC for constrained systems.

\section{Sampling from Constrained Systems and Hypothesis Testing}

Sampling from highly constrained systems and combinatorial calculations 
are discussed here as a variation of the theme of rare event sampling. 
Exact tests and data surrogation are introduced as an
application field of this idea, where efficient sampling from constrained 
systems is essential. 

For general issues on Monte Carlo approximate counting,
see \cite{jerrum1996markov},
\cite{RubinsteinKroese200712}, and \cite{Rubino_Tuffin200905}.

\subsection{MCMC Sampling from Constrained Systems} \label{sec:const}

MCMC sampling is difficult when constraints 
exist among \FFFRed{random} variables.
In such cases, it is often not easy to find 
a set of Metropolis--Hastings moves that realizes an ergodic Markov chain without violating the constraints. For example, considerable
effort is devoted to find ergodic moves for contingency tables with 
fixed margins and other constraints
(\FFRed{\cite{diaconis1998algebraic, bunea_besag2000, takemura2004some}})~\footnote{
See also \cite{jacobson1996generating} for an algorithm 
specialized for Latin squares;
it partially utilized a soft constraint strategy.
}. 
Although partial success has been obtained using highly sophisticated 
mathematics, the problem becomes increasingly difficult when 
problem complexity increases. 

Yet another general strategy for dealing with highly constrained systems is
an introduction of ``soft constraints.'' \RRRed{First, given constraints 
$f_i(x)=0$, $i=1, \cdots L$,  
we define statistics $\xi$ of the state variables $X$ that
satisfy the following conditions: (1)~$\xi(X) \geq 0$ and
(2)~$\xi(X) = 0$, if and only if $X$ satisfy $f_i(X)=0$ for all $i$.
A simple example of such statistics is  
\begin{equation*}
\xi(x)= \sum_{i=1}^L c_i | f_i(x) |^\alpha.
\end{equation*} }
Here, $\alpha >0$ and $c_i>0$ are arbitrary constants; 
 $\alpha=1$ is usually better than $\alpha=2$ because
$\xi$ keeps small values when $|f_i(x)|$ increases in the case of $\alpha=1$. 
Then, a finite 
value of $\xi$ represents {\it soft} constraints, whereas $\xi=0$
corresponds to the original {\it hard} constraints. 
\RRRed{Random sampling
of the value of $X$ usually gives a large value of $\xi(X)$; hence, $\xi(X)=0$ can be regarded as a ``rare event.''}

\RRRed{
At this point, we introduce
multicanonical MCMC with target statistics $\xi$ and sample
rare events $X$ defined by $\xi(X)=0$ (or, for a continuous variable $X$, 
$\xi(X) \approx 0$). Then, after tuning weights 
with the Wang--Landau algorithm, a production run 
provides samples of $X$ that (nearly) satisfy  
the constraints $f_i(X)=0$ (or $f_i(X) \approx 0$) 
for all $i$.} 
Note that a similar strategy can be implemented using
a combination of an exponential family with sufficient
statistics $\xi$ and replica
exchange MCMC;  in this case, a large value of $\beta$ 
corresponds to hard constraints. 

Some references are as follows~\footnote{
``Self-avoidingness'' of random walk is also
well treated by the soft constraint strategy discussed here;
see \cite{vorontsov1996free, vorontsov2004entropic,
iba1998simulation, chikenji1999multi,
2012arXiv1212.2181S}.
}. 
\cite{pinn1998number} introduced replica
exchange MCMC with soft constraints to this field, and 
the number of magic squares
of size $6\times 6$ is estimated in their paper. 
Kitajima and Kikuchi (private communication) 
extended it to $30\times 30$ using multicanonical 
MCMC. 
\cite{hukushima2002extended} estimated the number 
of N-queen configurations by replica exchange MCMC, while
\cite{zhang2009counting} treated N-queen and Latin squares 
using a hybrid of simulated
tempering (Sec.~\ref{sec:simt}) and the Wang--Landau 
algorithm; they dealt with Latin squares up to 
size $100\times 100$.  \cite{fishman2012counting}
proposed an approach based on soft constraints  
for counting contingency tables;  conventional MCMC is 
used in his paper.  

\subsection{Application to Hypothesis Testing} \label{sec:surr}

Here, we discuss how multicanonical MCMC 
(and also replica exchange MCMC) can
be useful for exact tests and data surrogation; the proposed method
is tested with a simple example of time series.

\subsubsection{MCMC Exact Tests}

MCMC is useful for implementing 
statistical tests with a complicated null distribution.
Particularly important cases occur when the null distribution 
is a distribution
conditioned with a set of statistics $\zeta_i$. In these cases,
the null hypothesis  is represented as the uniform
distribution of $X$ on the set defined by
$\zeta_i (x)=\zeta_i^{o}, \, i=1, \ldots L$,
where $X$ is a random variable and $\zeta_i^{o}$ 
is the value of statistics $\zeta_i$
corresponding to the observed data. For a continuous variable $X$,
this condition can be relaxed as
\begin{equation} \label{zeta2}
|\zeta_i (x) - \zeta_i^{o}| <\epsilon_i , \, i=1, \ldots L,
\end{equation}
where $\epsilon_i$ is a constant with a small value.

A prototype of such a test is Fisher's exact test of contingency
tables (\cite{agresti1992survey}), 
where the marginals of the table correspond to $\zeta_i$'s;
a number of extended versions exist
and MCMC
algorithms with complicated Metropolis moves have been developed for them, as mentioned in the previous section. \cite{besag1989generalized} described a test where an Ising model on the square lattice represents the null hypothesis. 

In our view, it is natural to
introduce the ``soft constraint'' strategy described 
in  Sec.~\ref{sec:const} to this problem.
\RRRed{When we define the statistics $\xi$ as 
$\xi(x)= \sum_{i=1}^L  |\zeta_i (x) - \zeta_i^{o} |$,
it is straightforward to apply multicanonical MCMC for sampling $X$ that
uniformly distributed on the set 
defined by  $\zeta_i (x)=\zeta_i^{o}, \, i=1, \ldots L$ or its
generalization \eqref{zeta2}.} This strategy is quite general 
and can be applied to a variety of MCMC hypothesis testing~\footnote{
As mentioned in the previous section,
\cite{yu2011efficient, LiangLiuCarroll201009} also discussed hypothesis testing
with stochastic approximation Monte Carlo, which can be regarded
as a version of multicanonical MCMC in this case.
They, however, focused on
the problem of calculating small p-values; it differs from our
idea of using multicanonical MCMC as a sampler 
from highly constrained systems.}. 

\subsubsection{Data Surrogation} \label{surr}

In  nonlinear dynamics and neural science,  
statistical tests for time series
based on \eqref{zeta2} are well 
developed (\cite{schreiber2000surrogate}).
They are called as {\it surrogate data} methods, 
and samples from null distributions 
defined by \eqref{zeta2} are called as {\it surrogates } 
of the original data. An example of the problem where surrogation is intensively used is testing of statistical properties of neural spike trains (\cite{Gruen_Rotter201008}). 

In conventional approaches, surrogates are generated by
partial randomization of the original data. For example, 
if the phase of time series data $x^o(t), t=1,\ldots, N$ 
is randomized after the
complex Fourier transform, then its inverse transform $x=\{x(t)\}$ 
has the same sets of 
correlation functions
\begin{equation}\label{C}
C(x; \tau)=\sum_{t=1}^{N-\tau} x(t) x(t+\tau) 
\end{equation}
as the original time series~\footnote{\FFFRed{To be precise,
we should assume a periodic boundary condition and change
the upper limit of the summation from $N-\tau$
to $N$.}} and is considered as 
a surrogate that maintains the value of sufficient statistics $\zeta_\tau(x)=C(x; \tau)$. Although a quick solution 
is provided in this case, 
solutions to general cases are only found on a case-by-case basis, and it becomes increasingly difficult as the complexity of the problems increases.

Therefore, Schreiber 
proposed a general idea of regarding data surrogation  as an optimization
problem (\cite{schreiber1998constrained, schreiber2000surrogate}).  
According to this idea, generating a surrogate is equivalent
to finding a solution of \eqref{zeta2}, which can be treated by
a general-purpose optimization algorithm, for example, simulated annealing.
An application of this idea in neural science is found
 in \cite{hirata2008testing}.

This was an epoch-making idea in this field;
randomization via a clever idea was no longer required, being 
replaced by a routine procedure at the cost of computational time. 
However, in data surrogation, we want to generate a sample 
(or a set of samples) unbiasedly selected from 
the null distribution defined by \eqref{zeta2}, 
and not obtain a sample that satisfies \eqref{zeta2}.

Therefore, applying multicanonical MCMC seems a better 
choice. Hence, we again arrive 
at the idea of exact testing with multicanonical MCMC. 

\subsubsection{Example} \label{timeseries}

\begin{figure}[tb]
\begin{center}
\begin{minipage}{4.5cm}
\begin{center}
\includegraphics[width=4cm]{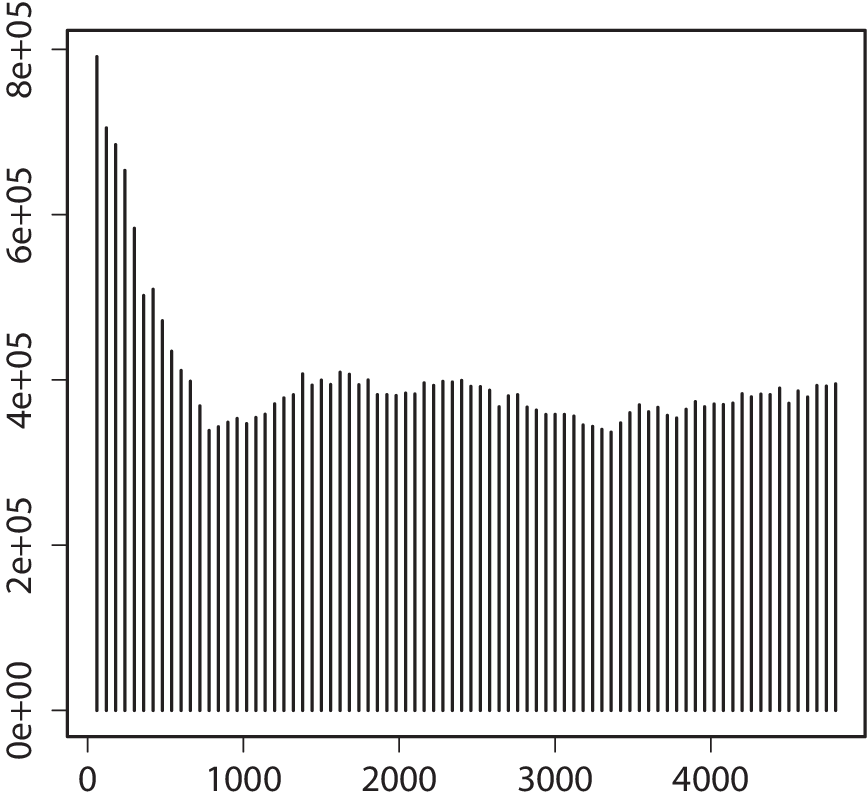}
\end{center}
\end{minipage}
\begin{minipage}{4.5cm}
\begin{center}
\includegraphics[width=4cm]{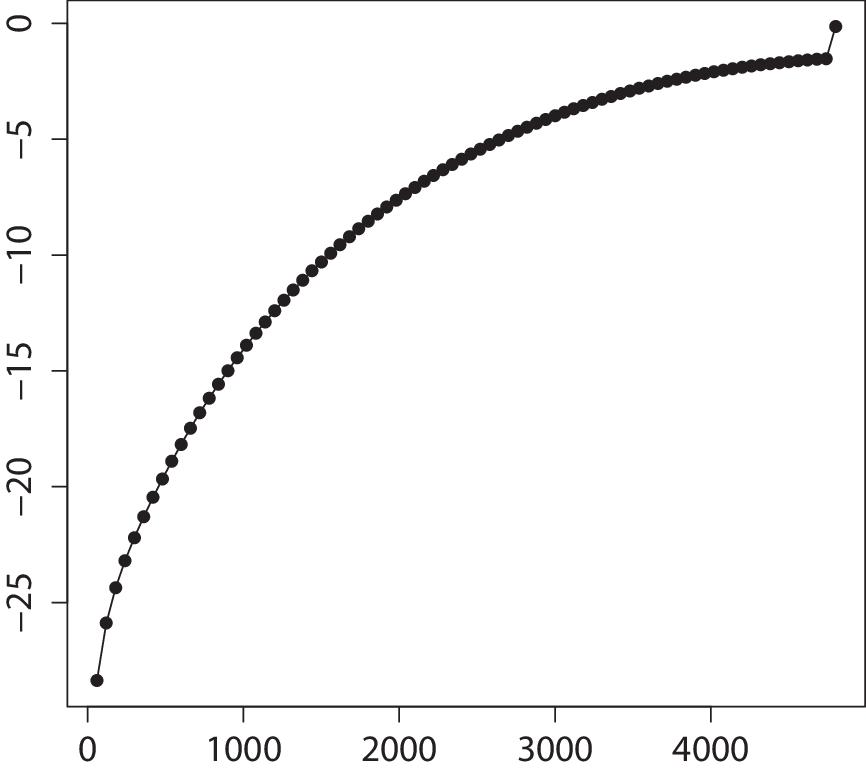}
\end{center}
\end{minipage}
\end{center}
\caption{\RRed{Left frequency of the occurrence of $\xi$ 
in the production run  of
multicanonical MCMC. The horizontal and vertical axes correspond to
the value of $\xi$ and the observed frequency 
in the given bins, respectively.
Right probability density of $\xi$.
The horizontal and vertical axes correspond to the value of $\xi$ and 
the estimated log-probabilities \FFFRed{($\log_{10}$)}, respectively;  a set of bins used in 
the left panel is also applied in the right panel 
for defining probabilities. 
The spike in the rightmost bin corresponds to a 
cumulated probability of  larger values of $\xi$.} 
[from Y.~Iba, IEICE Technical Report 
IBISML2011-7(2011-06), 43-50,
in Japanese, \copyright IEICE 2011]
}\label{fig:muls1}
\end{figure}

\begin{figure}[htb]
\begin{center}
\begin{minipage}{6.5cm}
\begin{center}
\includegraphics[width=6.3cm]{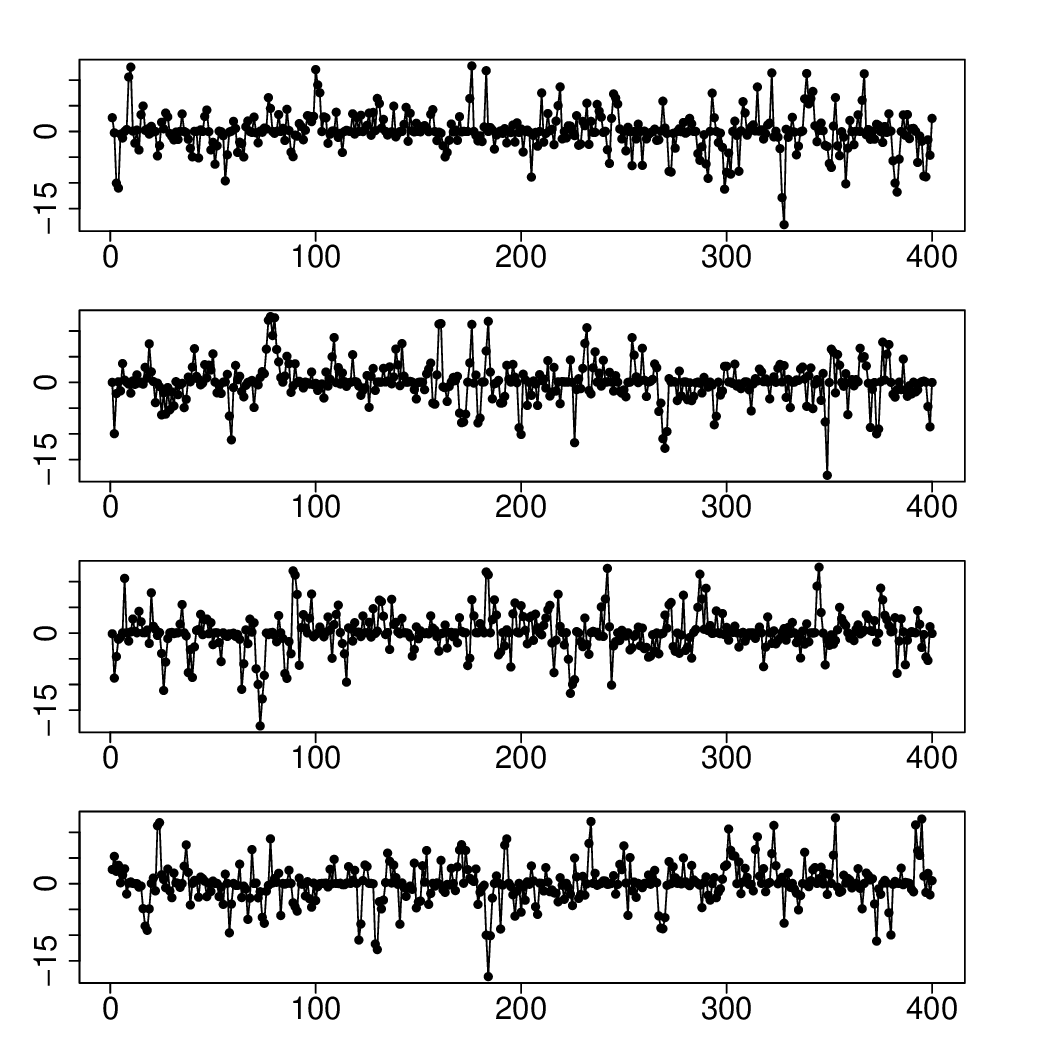}
\end{center}
\end{minipage}
\hspace{0.8cm}
\begin{minipage}{4cm}
\begin{center}
\includegraphics[width=4cm]{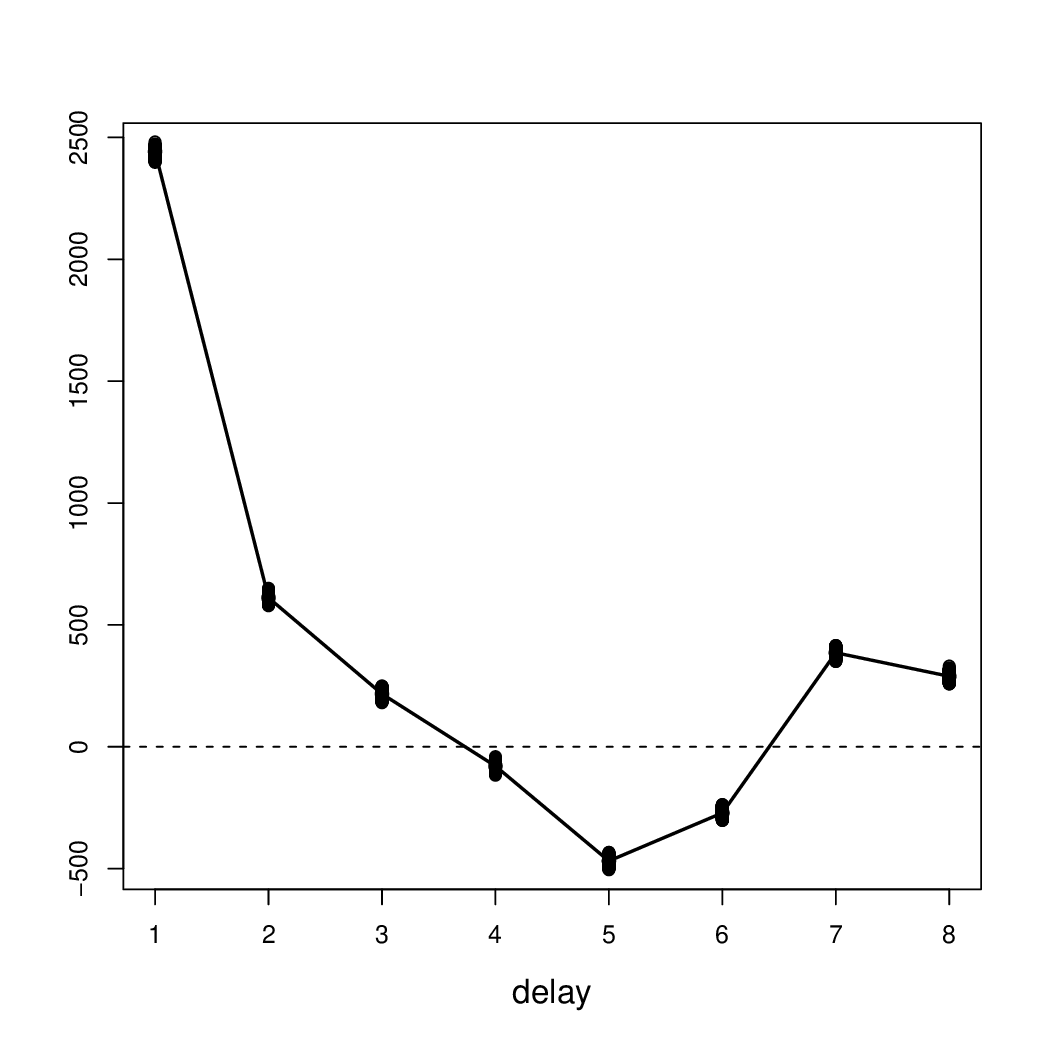}
\end{center}
\end{minipage}
\caption{\RRed{Left surrogate data generated by the proposed method. 
Uppermost series correspond to the original 
data and other three are surrogates. 
The horizontal and vertical axes correspond 
to $t$ and $x_t$, respectively. 
Right comparison of correlation functions. 
The horizontal and vertical axes correspond to the delay $\tau$ and 
the values of the correlation function, respectively.
The line represents $C(x^o, \tau)$, which corresponds to the original data, while black dots represent sets of $C(x^{(k)}, \tau)$ obtained from surrogated data; \IBlue{$N=400$} is the length of the time series.
The results of \IBlue{1976} samples are printed over each other; 
hence, symbols are almost overlapping.}
[from Y.~Iba, IEICE Technical Report 
IBISML2011-7(2011-06), 43-50, 
in Japanese, \copyright IEICE 2011]
}
\label{fig:muls2}
\end{center}
\end{figure}

Let us illustrate the idea of ``multicanonical surrogation'' using
an  example from \cite{schreiber1998constrained}~\footnote{
The results in this subsection (including Figs.~\ref{fig:muls1} 
and \ref{fig:muls2})
appeared in an IEICE Technical Report IBISML2011-7(2011-06) 
in Japanese, as a report without peer review. These have 
never been published in English.
}~\footnote{ 
\FFFRed{A quick practical solution is present for
this problem, but it is not a perfect one;
see \cite{schreiber1998constrained}.}
}. 
In this example, the problem is to generate
artificial time series $x=\{x_1,x_2,\ldots, x_N\}$
by permuting the original time series 
$x^o=\{x^o_1,x^o_2,\ldots, x^o_N\}$ given as observed data.
The constraint is to maintain the 
correlation functions $C(x;\tau)$, defined
as \eqref{C}, to be nearly equal to the original correlation functions 
$C(x^o;\tau)$ for $\tau=1 \ldots T$; here, the constant $0<T<N$ is 
the maximum of the delay $\tau$, where we
expect correlation coincidence. 

\RRed{
Here, 
\IBlue{$
\xi(X)=
\sum_{\tau=1}^T  \left |C(X;  \tau)-C(x^o; \tau) \right |
$}
is used to define multicanonical MCMC that samples
\FFFRed{$X=\{X_1,X_2, \cdots X_N\}$}.
$\xi(X)$ is zero if and only if $C(X; \tau)=C(x^o; \tau)$ for all
$1 \leq \tau \leq T$. Then, Metropolis--Hastings moves
are defined by the swap of a randomly selected pair. In detail, a pair
$i$ and $j$ is selected by a random number in 
each step and a new candidate $x^{\rm new}$ of $X$
is generated by $x^{\rm new}_i=x_j$ and  $x^{\rm new}_j=x_i$
without changing other components,
using the current values $X_i=x_i$ and $X_j=x_j$.
Here, the value of $\{X_i\}$ is initialized as a random permutation of
$\{x_i^o\}$. }

In the following experiment, 
we consider time series $x^o$ 
of length \IBlue{$N=400$} generated by nonlinear observations 
of a linear AR process $y$ driven by uniform noise, that is, 
$$
x_{t}^{o}=y_t^3, \,\,\,
y_{t+1}=0.3 y_t + \eta_t,  \,\,\, \eta_t \sim U(-2,2).
$$
Here, we choose $T=8$.
Multicanonical MCMC is designed for
realizing an approximately flat distribution
of $\xi$ in the interval \IBlue{$[0, 4800.8]$}, which is
divided into $\IBlue{80}$ bins~\footnote{Here, we round
the value of $\xi$ to $\xi^{\max}$ when 
it exceeds $\xi^{\max}$ instead of
rejecting the candidate; this causes the spike at  the right edge of the 
density in the right panel of Fig.~\ref{fig:muls1}. }.
\Red{In this choice of the interval, 
we consider two conditions: (1)~the interval contains
 a high entropy region where
the values of $\xi$ are readily realized by 
\FFFRed{a random permutation of the 
original time series}, and (2)~the last bin $\xi \simeq 0$
corresponds to a tail region of $\xi$ that we are interested in.}
The Wang--Landau
algorithm with \IBlue{$K_{\mathrm{max}}=15$} is used
to tune the weight; the $\sqrt{C}$ rule is utilized.  At each  
step of the iteration, we run MCMC until counts in each bin 
coincide with the value 
for the uniform histogram within \IBlue{1}\verb+%+ accuracy.  
The total number of Metropolis trials
is \IBlue{$2.5\times 10^8$}, of which \IBlue{$3.2\times 10^7$} 
are used for the final production run.

The results of this experiment are shown in Figs.~\ref{fig:muls1} and
\ref{fig:muls2}. In  Fig.~\ref{fig:muls1}, the distribution
of $\xi$ realized in the production run and the estimated log-density 
of $\xi$ are shown. The former is not quite
flat in a non-logarithmic scale, but enough 
to ensure efficient production of the desired samples. 
\RRRed{According to the right panel of Fig.~\ref{fig:muls1},
the probability of obtaining a sample within the bin 
$\xi \approx 0$ is estimated to be as small as \IBlue{$10^{-25}$ or less}, 
\FFFRed{assuming a random permutation of $x^o$.}}

In Fig.~\ref{fig:muls2}, the quality of the obtained samples 
is examined. In the left panel, three samples in the last bin 
$\xi \simeq 0$ are
shown, which are considerably different from one another. \RRRed{In the
right panel, correlation functions $C(x^{(k)}, \tau)$ are calculated for
each of the \IBlue{1976} samples $X=x^{(k)}$, $k=1, \ldots, \IBlue{1976}$,
in the bin $\xi \approx 0$ and compared to the original
$C(x^o, \tau)$, which indicate an extremely good 
agreement between them~\footnote{Note that
not all \IBlue{1976} samples are independent; 
some additional test is needed for estimating
the number of independent samples in our run. }.}

\section{Summary and Discussions}

In this paper, we discussed rare event sampling
using multicanonical MCMC. Two different methods
of tuning the  weight, entropic sampling and the Wang--Landau
algorithm, are explained. Then, examples for random matrices,
random graphs, chaotic dynamical systems, and 
data surrogation are shown. 
We hope our exposition will be useful for the exploration of
further novel applications of
multicanonical MCMC.





\appendix
\section{Appendix}

\subsection{Multicanonical MCMC for Exponential Family}

We begin this paper with a history of multicanonical MCMC;
it was originally developed as a method for sampling from 
Gibbs distributions, or an exponential family. Here, we
briefly discuss how to use multicanonical MCMC for
this original purpose.  

Assume that we want to compute the expectation 
$
{\mathbb E_\beta}[A(X)]=\sum_x A(x)\,\exp(\beta \xi(x))/Z_\beta
$
of statistics $A$
from the  output $X^{(i)}, i=1, \ldots, M $ obtained
from multicanonical MCMC  that realizes an almost  flat marginal
of $\xi$ in a ``sufficiently wide'' interval $[\xi_{\min}, \xi_{\max}]$. 
Then, for $M \rightarrow \infty$, the desired expectation is
computed by the {\it reweighting formula}~\footnote{
To use this formula for an off-line calculation of the average of $A$,
the values of $\xi$ and $A$ should be
recorded as pairs in the simulation, like 
$\mathbf{(} \, \xi(X^{(i)}), A(X^{(i)}) \, \mathbf{)}, 
\, i=1, \ldots, M$. 
}
\begin{equation} \label{eq:mul_aa}
{\mathbb E_\beta}[A(X)] \simeq 
\frac{\sum_{i=1}^M \left [ A(X^{(i)}) \,
 \tilde{P}(\xi (X^{(i)})) \,
\exp(\beta\xi(X^{(i)}) )\right ]}
{\sum_{i=1}^M 
\left [ \tilde{P}( \xi (X^{(i)} ) ) \,
\exp(\beta\xi(X^{(i)}) )\right ]}.
\end{equation}
\Red{
Further, we have an expression for the normalizing constant
$Z_\beta$ as
\begin{equation} \label{eq:mul_Z}
\frac{Z_\beta}{V} \simeq 
\frac{\sum_{i=1}^M \left [ 
 \tilde{P}(\xi (X^{(i)})) \,
\exp(\beta\xi(X^{(i)}) )\right ]}
{\sum_{i=1}^M 
\tilde{P}( \xi (X^{(i)} ) )},
\end{equation}
where $V$ is the total number of states
of \RRRed{the variable $X$ that satisfy $\xi_{\min}< \xi(X) <\xi_{\max}$};
it is useful for the calculation of marginal likelihood in 
statistics and free energy in physics.
}

\begin{figure}[bt]
\begin{center}
\includegraphics[width=5.5cm]{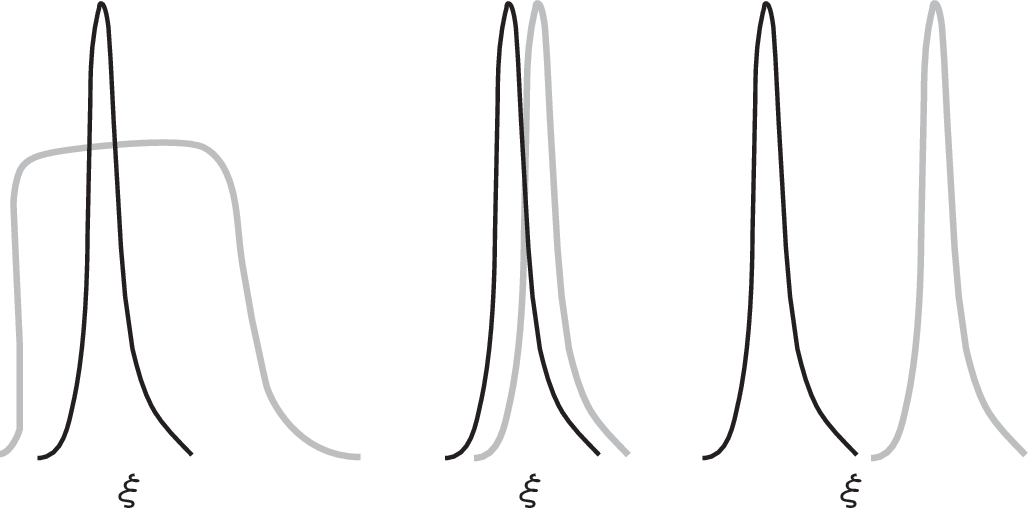}
\caption{Overlap of marginals of $\xi$. 
The horizontal axis corresponds to
the sufficient statistics $\xi$. Left a 
multicanonical weight (gray) and a member with a given $\beta$ 
of the exponential family (black).
Central and right panels  a pair of members with different values 
$\beta$ and $\beta^\prime$ of the
exponential family. The center panel corresponds to cases 
with a small $|\beta^\prime-\beta |$,
whereas the right panel corresponds to 
cases with a large $|\beta^\prime-\beta |$.  
  }\label{fig:overlap}
\end{center}
\end{figure}

It is easy to derive these expressions~\footnote{ 
Note that \eqref{eq:mul_aa} becomes
\eqref{eq:mul_a1}, if we substitute 
$I(\xi_0 \leq \xi(X^{(i)}))$ for $\exp(\beta\xi(X^{(i)}) )$.} considering 
that the
multicanonical weight is proportional to  $\tilde{P}( \xi (X^{(i)} ) )^{-1}$.
\
Expressions \eqref{eq:mul_aa} and \eqref{eq:mul_Z}, however, are quite unusual
in the sense that we can use them for a broad range of $\beta$ where
the interval $[\xi_{\min}, \xi_{\max}]$ covers a necessary region. Using
this property, multicanonical MCMC simultaneously gives the expectations
${\mathbb E_\beta}[A(X)]$
for all $\beta$, through a single production run of
a single chain. This is because
a multicanonical weight gives a flat distribution
of $\xi$ that has a considerable overlap with the distribution 
$\exp(\beta \xi(x))/Z_\beta$
for any value of $\beta$,  which is intuitively understood
from the left panel in Fig.~\ref{fig:overlap}. 

If we consider a similar
reweighing that uses outputs of MCMC at $\beta^\prime$ for computing 
the expectation at a different $\beta$, it is practically impossible
for a high-dimensional $X$ unless the difference $|\beta^\prime-\beta |$
is very small.  This is because the overlap of the distributions virtually vanishes as shown in the right panel of Fig.~\ref{fig:overlap};
in such cases, the variance of summands on the right-hand side of 
\eqref{eq:mul_aa} drastically increases.

\subsection{First-Order Transition and ``Phase Coexistence'' } \label{sec:co}

\begin{figure}[bt]
\begin{center}
\includegraphics[width=5cm]{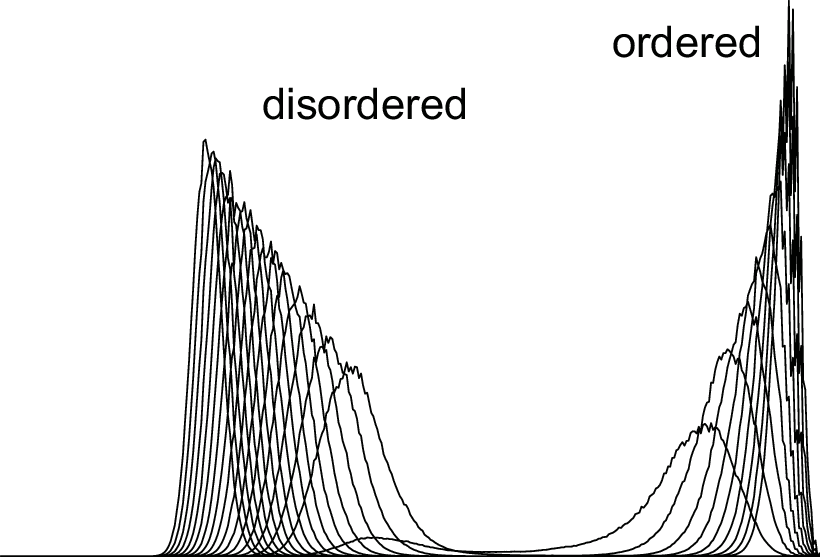}
\caption{
Marginals of $\xi$ with different values of $\beta$ in the case
of phase coexistence;
the horizontal axis corresponds to
the sufficient statistics $\xi$.
These curves are obtained from the 10-states Potts 
model (see \cite{berg1992multicanonical}), which consists
of discrete variables $\{X_i\}, \, X_i \in \{1, \cdots, 10\}$ 
on a square lattice; they are computed by reweighting of the
outputs of a single production run of multicanonical MCMC.  
If $\{X_i\}$ belongs to  an ``ordered'' component, most
$X_i$s take the same value. In contrast,
their values are almost random in the ``disordered'' component.   
In the disordered component,
$\xi$ takes smaller values, but the number of states of $X$ that
belongs to the component is large; hence, the total
probability is comparable in both components. }
\label{fig:multi2}
\end{center}
\end{figure}

As already mentioned in the main text, there are examples in which 
a region of $\xi$ is virtually not realized for any choice
of the canonical parameter $\beta$ of the exponential family 
with sufficient statistics $\xi$. The marginal distribution of $\xi$
has multiple peaks in this region of $\beta$, as  illustrated in Fig.~\ref{fig:multi2}.
Such examples naturally appear in statistical physics, when we study the
``phase coexistence'' phenomena near first-order 
phase transitions~\footnote{Ice 
and water coexist at 0 $^\circ$C; that is, both of them correspond to the same $\beta$ but
the values of average energy \Blue{$-{\mathbb E}(\xi)$} are different. }.
On the other hand, it seems that the significance of such 
phenomena in statistics and engineering
has not been fully explored.

In such cases, distributions defined by multicanonical weights
are not well approximated by a mixture of the members of the 
corresponding exponential family; this is easily understood by considering
Fig.~\ref{fig:multi2}. Hence, the advantage of replica exchange MCMC is 
limited because the sample path is blocked by the gap of $\xi$,
while multicanonical MCMC can, in principle, do better.
Both methods, however, seem to fail in very difficult cases; see 
\cite{iba2005exploration}.

\subsection{Simulated Tempering} \label{sec:simt}

\begin{figure}[bt]
\begin{center}
\includegraphics[width=7.8cm]{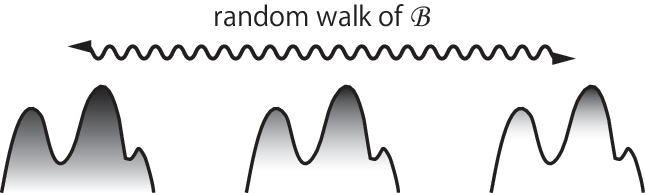}
\caption{\RRRed{Random walk of inverse temperature variable
$\mathpzc{B}$ in simulated tempering.
The vertical axis corresponds to the value of 
$\xi$, whereas the horizontal axis for each sub-chart
schematically represents a high-dimensional space of $X$.   
The values of inverse temperature
 $\beta$ are assumed to increase from left to right; 
the shading represents the corresponding changes 
in high-probability regions. 
Note that a random variable  $\mathpzc{B}$ is updated by MCMC, retaining
the value of $X$ at that time; in other words,
a separated procedure for changing $\mathpzc{B}$ is required for 
simulated tempering. }}\label{fig:tempering}
\end{center}
\end{figure}

The ``third'' method,
\FFFRed{{\it simulated tempering}} (\cite{marinari1992simulated, geyer1995annealing}), or {\it expanded ensemble Monte Carlo}
(\cite{lyubartsev1992new})~\footnote{This paper introduced an idea similar to
simulated tempering in an even more general 
framework.}, is briefly explained here. 
Practically, we recommend choosing
between multicanonical MCMC and
replica exchange MCMC. Simulated tempering, however, provides
an idea that interpolates these two algorithms and
is conceptually important.
\RRRed{The idea is simple---inverse temperature $\beta$
is regarded as a random variable (hereafter denoted by 
$\mathpzc{B}$) and we consider
MCMC sampling of $(X,\mathpzc{B})$ from \FFFRed{the combined distribution}
\begin{equation*}
P(x,\beta)=\frac{\exp(\beta \xi(x))}{Z_\beta} 
\, \pi(\beta)
= \exp(\beta \xi(x) -\log Z_\beta+\log \pi(\beta) ).
\end{equation*}
Hereafter, we choose a 
``pseudo prior'' $\pi(\beta)$ as a uniform density on $[\beta_{\min},
\beta_{\max}]$, resulting in a random walk of $\mathpzc{B}$ 
that uniformly covers the interval $[\beta_{\min}, \beta_{\max}]$; see
Fig.~\ref{fig:tempering}.
This behavior is similar to that of multicanonical MCMC, but
here $\mathpzc{B}$ is a variable updated in a separate step of 
MCMC; in contrast, a random walk 
of $\xi(X)$ is induced by the update of the state~$X$
in case of multicanonical MCMC.}

\RRRed{Although this concept is simple, a difficulty arises because
the MCMC update of $\mathpzc{B}$ requests the value of $Z_\beta$
as a function of $\beta$, which is unknown in most cases~\footnote{
The normalizing constant (partition function) 
$Z_\beta$ is not required for replica exchange MCMC, 
because it cancels in the Metropolis--Hastings ratio
necessary for deciding whether to 
accept/reject the swap of  the states between
chains; \FRed{this} is an
essential advantage of replica exchange MCMC.
}. 
Hence, we should introduce the estimation
of $Z_\beta$ using repeated preliminary runs, which is
similar to the weight tuning procedure in multicanonical MCMC.
See the above references, as well as 
\cite{zhang2007simulation}, which introduced
a method like the Wang--Landau algorithm.}

\subsection{Implementation on Parallel Hardware}\label{sec:pc}

Replica exchange MCMC is naturally parallelizable.
Then, how can we efficiently implement 
multicanonical MCMC on parallel hardware?
A simple solution is parallelization of the weight 
tuning stage. That is, a set of  preliminary runs
is performed in parallel, each of which
runs on a CPU;  they share
a histogram where total number of visits to each value
of $\xi$ is recorded. Some variants of this idea are discussed 
in, for example, \cite{zhan2008parallel, 2011arXiv1109.3829B}. 
If we want to go beyond these schemes, something
more intricate is required. For example, the range of 
$\xi$ is divided into a set of intervals and a multicanonical weight
is realized in each of them; see 
\cite{wang2001determining, Mitsutake01,PhysRevLett.110.210603}. 
\FFRed{In the latter two studies}, 
the exchange of states between neighboring intervals 
is incorporated.

\subsection{Multivariate Extensions}

Multicanonical MCMC samples a high-dimensional $X$, while
adaptation of the weight is performed in a one-dimensional space of $\xi$.
It is possible to introduce  a ``multivariate
multicanonical weight,'' which realizes an
almost uniform density in a region of  two-dimensional $(\xi_1, \xi_2)$ or even three-dimensional $(\xi_1, \xi_2, \xi_3)$ \RRed{spaces}, where $\xi_k$ is
a function of $X$.
Examples of such extensions are found in
\cite{shteto1997monte, iba1998simulation, 
higo1998two, chikenji1999multi, 
chikenji2000role, yan2002density, zhou2006wang}. 
Usually, the adaptation of weights in a multivariate case is
more difficult than in a univariate case, because of the sparseness of the
data collected in the preliminary run; \cite{zhou2006wang}
proposed the use of kernel density estimation for this problem.

\begin{acknowledgements}
The authors would like to thank Koji Hukushima for the helpful discussions
and the permission for the use of figures in \cite{saito2010multicanonical}.
\FRed{We would also be grateful to Arnaud Doucet and the referees, for
their helpful advice allowed us to improve
the manuscript. This work was supported by JSPS KAKENHI 
Grant Numbers 22500217, 25330299, and 25240036.}
Saito is supported by a Grant-in-Aid for Scientific Research (No.
21120004) on Innovative Areasg Neural creativity for communication
h (No. 4103), and the Platform for Dynamic Approaches to
Living System from MEXT, Japan.
\end{acknowledgements}


\begin{thebibliography}{119}
\providecommand{\natexlab}[1]{#1}
\providecommand{\url}[1]{{#1}}
\providecommand{\urlprefix}{URL }
\expandafter\ifx\csname urlstyle\endcsname\relax
  \providecommand{\doi}[1]{DOI~\discretionary{}{}{}#1}\else
  \providecommand{\doi}{DOI~\discretionary{}{}{}\begingroup
  \urlstyle{rm}\Url}\fi
\providecommand{\eprint}[2][]{\url{#2}}

\bibitem[{Aazami and Easther(2006)}]{aazami2006cosmology}
Aazami A, Easther R (2006) {Cosmology from random multifield potentials}.
  Journal of Cosmology and Astroparticle Physics 3(03):013

\bibitem[{Agresti(1992)}]{agresti1992survey}
Agresti A (1992) A survey of exact inference for contingency tables.
  Statistical Science 7(1):131--153

\bibitem[{Atchad{\'e} and Liu(2010)}]{atchade2010wang}
Atchad{\'e} YF, Liu JS (2010) The {Wang--Landau} algorithm in general state
  spaces: applications and convergence analysis. Statistica Sinica
  20(1):209--233

\bibitem[{Bachmann and Janke(2003)}]{bachmann2003multicanonical}
Bachmann M, Janke W (2003) Multicanonical chain-growth algorithm. Physical
  Review Letters 91(20):208105

\bibitem[{Baumann(1987)}]{baumann1987noncanonical}
Baumann B (1987) Noncanonical path and surface simulation. Nuclear Physics B
  285:391--409

\bibitem[{Beck and Schl{\"o}gl(1993)}]{BeckSchloegl199501}
Beck C, Schl{\"o}gl F (1993) Thermodynamics of Chaotic Systems: An
  Introduction. Cambridge University Press, Cambridge

\bibitem[{Belardinelli and Pereyra(2007{\natexlab{a}})}]{PhysRevE.75.046701}
Belardinelli R, Pereyra V (2007{\natexlab{a}}) Fast algorithm to calculate
  density of states. Physical Review E 75:046701

\bibitem[{Belardinelli and Pereyra(2007{\natexlab{b}})}]{belardinelli2007wang}
Belardinelli R, Pereyra V (2007{\natexlab{b}}) {Wang--Landau} algorithm: A
  theoretical analysis of the saturation of the error. The Journal of Chemical
  Physics 127:184105

\bibitem[{Berg(2000)}]{berg2000introduction}
Berg BA (2000) Introduction to multicanonical {Monte Carlo} simulations. Fields
  Institute Communications 26:1--24

\bibitem[{Berg(2004)}]{Berg200410}
Berg BA (2004) {Markov} Chain {Monte Carlo} Simulations and Their Statistical
  Analysis. World Scientific, Singapore

\bibitem[{Berg and Celik(1992)}]{berg1992new}
Berg BA, Celik T (1992) New approach to spin-glass simulations. Physical Review
  Letters 69(15):2292--2295

\bibitem[{Berg and Neuhaus(1991)}]{berg1991multicanonical}
Berg BA, Neuhaus T (1991) Multicanonical algorithms for first order phase
  transitions. Physics Letters B 267(2):249--253

\bibitem[{Berg and Neuhaus(1992)}]{berg1992multicanonical}
Berg BA, Neuhaus T (1992) Multicanonical ensemble: A new approach to simulate
  first-order phase transitions. Physical Review Letters 68(1):9--12

\bibitem[{Besag and Clifford(1989)}]{besag1989generalized}
Besag J, Clifford P (1989) Generalized {Monte Carlo} significance tests.
  Biometrika 76(4):633--642

\bibitem[{Binder and Heermann(2012)}]{BinderHeermann201210}
Binder K, Heermann D (2012) {Monte Carlo} Simulation in Statistical Physics: An
  Introduction. Springer, Berlin

\bibitem[{{Birge} et~al(2012){Birge}, {Chang}, and
  {Polson}}]{2012arXiv1212.0534B}
{Birge} JR, {Chang} C, {Polson} NG (2012) Split sampling: Expectations,
  normalisation and rare events. ArXiv e-prints \eprint{1212.0534}

\bibitem[{Bononi et~al(2009)Bononi, Rusch, Ghazisaeidi, Vacondio, and
  Rossi}]{bononi2009fresh}
Bononi A, Rusch L, Ghazisaeidi A, Vacondio F, Rossi N (2009) A fresh look at
  multicanonical {Monte Carlo} from a telecom perspective. In: Global
  Telecommunications Conference, 2009. GLOBECOM 2009, IEEE, pp 1--8

\bibitem[{Bornn et~al(2013)Bornn, Jacob, Del~Moral, and
  Doucet}]{2011arXiv1109.3829B}
Bornn L, Jacob PE, Del~Moral P, Doucet A (2013) An adaptive interacting
  {Wang--Landau} algorithm for automatic density exploration. Journal of
  Computational and Graphical Statistics 22(3):749--773

\bibitem[{Botev et~al(2013)Botev, L'Ecuyer, and Tuffin}]{Botev2013}
Botev ZI, L'Ecuyer P, Tuffin B (2013) {Markov} chain importance sampling with
  applications to rare event probability estimation. Statistics and Computing
  23(2):271--285

\bibitem[{Brooks et~al(2011)Brooks, Gelman, Jones, and
  Meng}]{Brooks_Gelman_Jones_Meng201105}
Brooks S, Gelman A, Jones GL, Meng XL (eds)  (2011) Handbook of {Markov} Chain
  {Monte Carlo}. Chapman and Hall/CRC, New York

\bibitem[{Bucklew(2004)}]{Bucklew200403}
Bucklew JA (2004) Introduction to Rare Event Simulation (Springer Series in
  Statistics). Springer, New York

\bibitem[{Bunea and Besag(2000)}]{bunea_besag2000}
Bunea F, Besag J (2000) {MCMC} in ${I} \times {J} \times {K}$ contingency
  tables. Fields Institute Communications 26:25--36

\bibitem[{Calvo(2002)}]{calvo2002sampling}
Calvo F (2002) Sampling along reaction coordinates with the {Wang--Landau}
  method. Molecular Physics 100(21):3421--3427

\bibitem[{Chikenji and Kikuchi(2000)}]{chikenji2000role}
Chikenji G, Kikuchi M (2000) What is the role of non-native intermediates of
  $\beta$-lactoglobulin in protein folding? Proceedings of the National Academy
  of Sciences 97(26):14,273--14,277

\bibitem[{Chikenji et~al(1999)Chikenji, Kikuchi, and Iba}]{chikenji1999multi}
Chikenji G, Kikuchi M, Iba Y (1999) Multi-self-overlap ensemble for protein
  folding: ground state search and thermodynamics. Physical Review Letters
  83(9):1886--1889

\bibitem[{Chopin et~al(2012)Chopin, Leli{\`e}vre, and Stoltz}]{chopin2012free}
Chopin N, Leli{\`e}vre T, Stoltz G (2012) Free energy methods for {Bayesian}
  inference: efficient exploration of univariate {Gaussian} mixture posteriors.
  Statistics and Computing 22(4):897--916

\bibitem[{Dean and Majumdar(2008)}]{dean2008extreme}
Dean DS, Majumdar SN (2008) {Extreme value statistics of eigenvalues of
  Gaussian random matrices}. Physical Review E 77(4):041108

\bibitem[{de~Oliveira et~al(1998)de~Oliveira, Penna, and
  Herrmann}]{OLIVEIRA1998}
de~Oliveira PMC, Penna TJP, Herrmann HJ (1998) Broad histogram {Monte Carlo}.
  The European Physical Journal B - Condensed Matter and Complex Systems
  1(2):205--208

\bibitem[{Diaconis and Sturmfels(1998)}]{diaconis1998algebraic}
Diaconis P, Sturmfels B (1998) Algebraic algorithms for sampling from
  conditional distributions. The Annals of statistics 26(1):363--397

\bibitem[{Donetti et~al(2005)Donetti, Hurtado, and
  Mu{\~n}oz}]{donetti2005entangled}
Donetti L, Hurtado PI, Mu{\~n}oz MA (2005) Entangled networks, synchronization,
  and optimal network topology. Physical Review Letters 95(18):188701

\bibitem[{Donetti et~al(2006)Donetti, Neri, and Mu{\~n}oz}]{donetti2006optimal}
Donetti L, Neri F, Mu{\~n}oz MA (2006) Optimal network topologies: Expanders,
  cages, {Ramanujan} graphs, entangled networks and all that. Journal of
  Statistical Mechanics: Theory and Experiment 2006(08):P08007

\bibitem[{Driscoll and Maki(2007)}]{driscoll2007searching}
Driscoll TA, Maki KL (2007) Searching for rare growth factors using
  multicanonical {Monte Carlo} methods. SIAM Review 49(4):673--692

\bibitem[{Fishman(2012)}]{fishman2012counting}
Fishman GS (2012) Counting contingency tables via multistage {Markov} chain
  {Monte Carlo}. Journal of Computational and Graphical Statistics
  21(3):713--738

\bibitem[{{Fort} et~al(2012){Fort}, {Jourdain}, {Kuhn}, {Leli{\`e}vre}, and
  {Stoltz}}]{2012arXiv1207.6880F}
{Fort} G, {Jourdain} B, {Kuhn} E, {Leli{\`e}vre} T, {Stoltz} G (2012)
  Convergence and efficiency of the {Wang--Landau} algorithm. ArXiv e-prints
  \eprint{1207.6880}

\bibitem[{Frenkel and Smit(2002)}]{FrenkelSmit200111}
Frenkel D, Smit B (2002) Understanding Molecular Simulation, From Algorithms to
  Applications (Computational Science Series), 2nd edn. Academic Press, San
  Diego

\bibitem[{Geiger and Dellago(2010)}]{geiger2010identifying}
Geiger P, Dellago C (2010) Identifying rare chaotic and regular trajectories in
  dynamical systems with {Lyapunov} weighted path sampling. Chemical Physics
  375(2-3):309--315

\bibitem[{Geyer(1991)}]{citeulike:606345}
Geyer CJ (1991) {Markov} chain {Monte Carlo} maximum likelihood. In: Keramidas
  E (ed) Computing science and statistics: Proceedings of 23rd Symposium on the
  Interface, Interface Foundation, Fairfax Station, pp 156--163

\bibitem[{Geyer and Thompson(1995)}]{geyer1995annealing}
Geyer CJ, Thompson EA (1995) Annealing {Markov} chain {Monte Carlo} with
  applications to ancestral inference. Journal of the American Statistical
  Association 90(431):909--920

\bibitem[{Gilks et~al(1996)Gilks, Richardson, and
  Spiegelhalter}]{Gilks_Richardson_Spiegelhalter199512}
Gilks WR, Richardson S, Spiegelhalter DJ (eds)  (1996) {Markov} Chain {Monte
  Carlo} in Practice. Chapman and Hall, London

\bibitem[{Gr\"{u}n and Rotter(2010)}]{Gruen_Rotter201008}
Gr\"{u}n S, Rotter S (eds)  (2010) Analysis of Parallel Spike Trains (Springer
  Series in Computational Neuroscience). Springer, New York

\bibitem[{Hartmann(2002)}]{hartmann2002sampling}
Hartmann AK (2002) {Sampling rare events: statistics of local sequence
  alignments}. Physical Review E 65(5):056102

\bibitem[{Higo et~al(1997)Higo, Nakajima, Shirai, Kidera, and
  Nakamura}]{higo1998two}
Higo J, Nakajima N, Shirai H, Kidera A, Nakamura H (1997) Two-component
  multicanonical {Monte Carlo} method for effective conformation sampling.
  Journal of computational chemistry 18(16):2086--2092

\bibitem[{Higo et~al(2012)Higo, Ikebe, Kamiya, and Nakamura}]{higo2012enhanced}
Higo J, Ikebe J, Kamiya N, Nakamura H (2012) Enhanced and effective
  conformational sampling of protein molecular systems for their free energy
  landscapes. Biophysical Reviews 4:27--44

\bibitem[{Hirata et~al(2008)Hirata, Katori, Shimokawa, Suzuki, Blenkinsop,
  Lang, and Aihara}]{hirata2008testing}
Hirata Y, Katori Y, Shimokawa H, Suzuki H, Blenkinsop TA, Lang EJ, Aihara K
  (2008) Testing a neural coding hypothesis using surrogate data. Journal of
  Neuroscience Methods 172(2):312--322

\bibitem[{Holzl{\"{o}}hner and Menyuk(2003)}]{holzloohner2003use}
Holzl{\"{o}}hner R, Menyuk CR (2003) {Use of multicanonical Monte Carlo
  simulations to obtain accurate bit error rates in optical communications
  systems}. Optics Letters 28(20):1894--1896

\bibitem[{Holzl{\"{o}}hner et~al(2005)Holzl{\"{o}}hner, Mahadevan, Menyuk,
  Morris, and Zweck}]{holzlohner2005evaluation}
Holzl{\"{o}}hner R, Mahadevan A, Menyuk CR, Morris JM, Zweck J (2005)
  {Evaluation of the very low BER of FEC codes using dual adaptive importance
  sampling}. IEEE Communications Letters 9(2):163--165

\bibitem[{Hukushima(2002)}]{hukushima2002extended}
Hukushima K (2002) Extended ensemble {Monte Carlo} approach to hardly relaxing
  problems. Computer Physics Communications 147(1--2):77--82

\bibitem[{Hukushima and Iba(2008)}]{hukushima2008monte}
Hukushima K, Iba Y (2008) {A Monte Carlo algorithm for sampling rare events:
  application to a search for the Griffiths singularity}. Journal of Physics:
  Conference Series 95:012005

\bibitem[{Hukushima and Nemoto(1996)}]{HUKUSHIMAKoji1996}
Hukushima K, Nemoto K (1996) {Exchange Monte Carlo method and application to
  spin glass simulations}. Journal of the Physical Society of Japan
  65(6):1604--1608

\bibitem[{Iba(2001)}]{iba2001extended}
Iba Y (2001) {Extended ensemble Monte Carlo}. International Journal of Modern
  Physics C 12(05):623--656

\bibitem[{Iba and Hukushima(2008)}]{yukito2008testing}
Iba Y, Hukushima K (2008) Testing error correcting codes by multicanonical
  sampling of rare events. Journal of the Physical Society of Japan
  77(10):103801

\bibitem[{Iba and Takahashi(2005)}]{iba2005exploration}
Iba Y, Takahashi H (2005) Exploration of multi-dimensional density of states by
  multicanonical {Monte Carlo} algorithm. Progress of Theoretical Physics
  Supplements 157:345--348

\bibitem[{Iba et~al(1998)Iba, Chikenji, and Kikuchi}]{iba1998simulation}
Iba Y, Chikenji G, Kikuchi M (1998) Simulation of lattice polymers with
  multi-self-overlap ensemble. Journal of the Physical Society of Japan
  67:3327--3330

\bibitem[{{Jacob} and {Ryder}(2011)}]{2011arXiv1110.4025J}
{Jacob} PE, {Ryder} RJ (2011) The {Wang--Landau} algorithm reaches the flat
  histogram criterion in finite time. ArXiv e-prints \eprint{1110.4025}

\bibitem[{Jacobson and Matthews(1996)}]{jacobson1996generating}
Jacobson MT, Matthews P (1996) Generating uniformly distributed random {Latin}
  squares. Journal of Combinatorial Designs 4(6):405--437

\bibitem[{Janke(1998)}]{janke1998multicanonical}
Janke W (1998) Multicanonical {Monte Carlo} simulations. Physica A: Statistical
  Mechanics and its Applications 254(1-2):164--178

\bibitem[{Jerrum and Sinclair(1996)}]{jerrum1996markov}
Jerrum M, Sinclair A (1996) The {Markov} chain {Monte Carlo} method: an
  approach to approximate counting and integration. Approximation algorithms
  for NP-hard problems pp 482--520

\bibitem[{Kastner et~al({2013})Kastner, Braumann, Man, Mosbach, Brownbridge,
  Akroyd, Kraft, and Himawan}]{Kastner2013244}
Kastner CA, Braumann A, Man PLW, Mosbach S, Brownbridge GPE, Akroyd J, Kraft M,
  Himawan C ({2013}) {Bayesian} parameter estimation for a jet-milling model
  using {Metropolis-Hastings} and {Wang--Landau} sampling. {Chemical
  Engineering Science} {89}:{244 -- 257}

\bibitem[{Kimura and Taki(1991)}]{2074066}
Kimura K, Taki K (1991) {Time-homogeneous parallel annealing algorithm}.
  {Proceedings of the 13th IMACS World Congress on Computation and Applied
  Mathematics (IMACS'91)} 2:827--828

\bibitem[{Kirkpatrick et~al(1983)Kirkpatrick, Gelatt, and
  Vecchi}]{kirkpatrick1983optimization}
Kirkpatrick S, Gelatt CD, Vecchi MP (1983) Optimization by simulated annealing.
  Science 220(4598):671--680

\bibitem[{Kitajima and Iba(2011)}]{KitajimaI11}
Kitajima A, Iba Y (2011) Multicanonical sampling of rare trajectories in
  chaotic dynamical systems. Computer Physics Communications 182(1):251--253

\bibitem[{K\"{o}rner et~al(2006)K\"{o}rner, Katzgraber, and
  Hartmann}]{koerner2006probing}
K\"{o}rner M, Katzgraber HG, Hartmann AK (2006) Probing tails of energy
  distributions using importance-sampling in the disorder with a guiding
  function. Journal of Statistical Mechanics: Theory and Experiment
  2006(04):P04005

\bibitem[{Kumar (2013)}]{Kumar2013}
Kumar S (2013). Random matrix ensembles: Wang-Landau algorithm
for spectral densities. Europhysics Letters, 101(2), 20002.

\bibitem[{Kwon and Lee(2008)}]{Kwon2008}
Kwon J, Lee KM (2008) Tracking of abrupt motion using {Wang--Landau Monte
  Carlo} estimation. In: Proceedings of the 10th European Conference on
  Computer Vision: Part I, Springer-Verlag, Berlin, Heidelberg, ECCV '08, pp
  387--400

\bibitem[{Laffargue et~al(2013)Laffargue, Lam, Kurchan, and
  Tailleur}]{2013arXiv1302.6254L}
Laffargue T, Lam KDNT, Kurchan J, Tailleur J (2013) Large deviations of
  {Lyapunov} exponents. Journal of Physics A: Mathematical and Theoretical
  46(25):254002

\bibitem[{Landau and Binder(2009)}]{LandauBinder201311}
Landau DP, Binder K (2009) A Guide to {Monte Carlo} Simulations in Statistical
  Physics, 3rd edn. Cambridge University Press

\bibitem[{Landau et~al(2004)Landau, Tsai, and Exler}]{landau2004new}
Landau DP, Tsai SH, Exler M (2004) A new approach to {Monte Carlo} simulations
  in statistical physics: {Wang--Landau} sampling. American Journal of Physics
  72(10):1294--1302

\bibitem[{Lee et~al(2006)Lee, Okabe, and Landau}]{lee2006convergence}
Lee HK, Okabe Y, Landau DP (2006) {Convergence and refinement of the
  Wang--Landau algorithm}. Computer Physics Communications 175(1):36--40


\bibitem[{Lee(1993)}]{lee1993new}
Lee J (1993) New {Monte Carlo} algorithm: entropic sampling.
  Physical Review Letters 71(2):211--214

\bibitem[{Liang(2005)}]{liang2005generalized}
Liang F (2005) A generalized {Wang--Landau} algorithm for {Monte Carlo}
  computation. Journal of the American Statistical Association
  100(472):1311--1327

\bibitem[{Liang et~al(2007)Liang, Liu, and Carroll}]{liang2007stochastic}
Liang F, Liu C, Carroll RJ (2007) Stochastic approximation in {Monte Carlo}
  computation. Journal of the American Statistical Association
  102(477):305--320

\bibitem[{Liang et~al(2010)Liang, Liu, and Carroll}]{LiangLiuCarroll201009}
Liang F, Liu C, Carroll RJ (2010) Advanced {Markov} Chain {Monte Carlo}
  Methods: Learning from Past Samples (Wiley Series in Computational
  Statistics). Wiley, West Sussex

\bibitem[{Lyubartsev et~al(1992)Lyubartsev, Martsinovski, Shevkunov, and
  Vorontsov-Velyaminov}]{lyubartsev1992new}
Lyubartsev AP, Martsinovski AA, Shevkunov SV, Vorontsov-Velyaminov PN (1992)
  {New approach to Monte Carlo calculation of the free energy: Method of
  expanded ensembles}. The Journal of Chemical Physics 96(3):1776--1783

\bibitem[{Majumdar and Vergassola(2009)}]{majumdar2009large}
Majumdar SN, Vergassola M (2009) {Large deviations of the maximum eigenvalue
  for Wishart and Gaussian random matrices}. Physical Review Letters
  102(6):060601

\bibitem[{Marinari and Parisi(1992)}]{marinari1992simulated}
Marinari E, Parisi G (1992) Simulated tempering: a new {Monte Carlo} scheme.
  Europhysics Letters 19(6):451--458

\bibitem[{Matsuda et~al(2008)Matsuda, Nishimori, and
  Hukushima}]{matsuda2008distribution}
Matsuda Y, Nishimori H, Hukushima K (2008) The distribution of {Lee--Yang}
  zeros and {Griffiths} singularities in the $\pm$ {J} model of spin glasses.
  Journal of Physics A: Mathematical and Theoretical 41(32):324012

\bibitem[{May(1972)}]{may}
May RM (1972) {Will a large complex system be stable?} Nature 238:413--414

\bibitem[{Mezei(1987)}]{mezei1987adaptive}
Mezei M (1987) Adaptive umbrella sampling: self-consistent determination of the
  {non-Boltzmann} bias. Journal of Computational Physics 68(1):237--248

\bibitem[{Mitsutake et~al(2001)Mitsutake, Sugita, and Okamoto}]{Mitsutake01}
Mitsutake A, Sugita Y, Okamoto Y (2001) Generalized-ensemble algorithms for
  molecular simulations of biopolymers. Biopolymers (Peptide Science)
  60(2):96--123

\bibitem[{Monthus and Garel(2006)}]{monthus2006probing}
Monthus C, Garel T (2006) Probing the tails of the ground-state energy
  distribution for the directed polymer in a random medium of dimension d= 1,
  2, 3 via a {Monte Carlo} procedure in the disorder. Physical Review E
  74(5):051109

\bibitem[{Newman and Barkema(1999)}]{newman1999monte}
Newman MEJ, Barkema GT (1999) {Monte Carlo} Methods in Statistical Physics.
  Clarendon Press, New York

\bibitem[{Ott(2002)}]{Ott02}
Ott E (2002) Chaos in Dynamical Systems. Cambridge University Press, Chambridge

\bibitem[{Pinn and Wieczerkowski(1998)}]{pinn1998number}
Pinn K, Wieczerkowski C (1998) Number of magic squares from parallel tempering
  {Monte Carlo}. International Journal of Modern Physics C 09(04):541--546

\bibitem[{Prellberg and Krawczyk(2004)}]{prellberg2004flat}
Prellberg T, Krawczyk J (2004) Flat histogram version of the pruned and
  enriched {Rosenbluth} method. Physical Review Letters 92(12):120602

\bibitem[{Robert and Casella(2004)}]{RobertCasella200508}
Robert CP, Casella G (2004) {Monte Carlo} Statistical Methods, 2nd edn.
  Springer, New York

\bibitem[{Rubino and Tuffin(2009)}]{Rubino_Tuffin200905}
Rubino G, Tuffin B (eds)  (2009) Rare Event Simulation using {Monte Carlo}
  Methods. Wiley, West Sussex

\bibitem[{Rubinstein and Kroese(2008)}]{RubinsteinKroese200712}
Rubinstein RY, Kroese DP (2008) Simulation and the {Monte Carlo} Method (Wiley
  Series in Probability and Statistics), 2nd edn. Wiley-Interscience, Hoboken

\bibitem[{Saito and Iba(2011)}]{SaitoI11}
Saito N, Iba Y (2011) Probability of graphs with large spectral gap by
  multicanonical {Monte Carlo}. Computer Physics Communications 182(1):223--225

\bibitem[{Saito et~al(2010)Saito, Iba, and Hukushima}]{saito2010multicanonical}
Saito N, Iba Y, Hukushima K (2010) Multicanonical sampling of rare events in
  random matrices. Physical Review E 82(3):031142

\bibitem[{Sasa and Hayashi(2006)}]{Sasa06}
Sasa S, Hayashi K (2006) Computation of the {Kolmogorov--Sinai} entropy using
  statistical mechanics: Application of an exchange {Monte Carlo} method.
  Europhysics Letters 74(1):156--162

\bibitem[{Schreiber(1998)}]{schreiber1998constrained}
Schreiber T (1998) Constrained randomization of time series data. Physical
  Review Letters 80(10):2105--2108

\bibitem[{Schreiber and Schmitz(2000)}]{schreiber2000surrogate}
Schreiber T, Schmitz A (2000) Surrogate time series. Physica D: Nonlinear
  Phenomena 142(3-4):346--382

\bibitem[{{Schulz} et~al(2003){Schulz}, {Binder}, {M{\"u}ller}, and
  {Landau}}]{2003PhRvE}
{Schulz} BJ, {Binder} K, {M{\"u}ller} M, {Landau} DP (2003) Avoiding boundary
  effects in {Wang--Landau} sampling. Physical Review E 67(6):067102

\bibitem[{Shell et~al(2002)Shell, Debenedetti, and
  Panagiotopoulos}]{shell2002generalization}
Shell MS, Debenedetti PG, Panagiotopoulos AZ (2002) Generalization of the
  {Wang--Landau} method for off-lattice simulations. Physical Review E
  66(5):056703

\bibitem[{Shirai and Kikuchi(2013)}]{2012arXiv1212.2181S}
Shirai NC, Kikuchi M (2013) Multicanonical simulation of the {Domb--Joyce}
  model and the {G\={o}} model: new enumeration methods for self-avoiding
  walks. Journal of Physics: Conference Series 454(1):012039

\bibitem[{Shteto et~al(1997)Shteto, Linares, and Varret}]{shteto1997monte}
Shteto I, Linares J, Varret F (1997) {Monte Carlo entropic sampling for the
  study of metastable states and relaxation paths}. Physical Review E
  56(5):5128--5137

\bibitem[{Sweet et~al(2001)Sweet, Nusse, and Yorke}]{Sweet01}
Sweet D, Nusse HE, Yorke JA (2001) Stagger-and-step method: Detecting and
  computing chaotic saddles in higher dimensions. Physical Review Letters
  86(11):2261--2264

\bibitem[{Tailleur and Kurchan(2007)}]{tailleur2007probing}
Tailleur J, Kurchan J (2007) Probing rare physical trajectories with {Lyapunov}
  weighted dynamics. Nature Physics 3(3):203--207

\bibitem[{Takemura and Aoki(2004)}]{takemura2004some}
Takemura A, Aoki S (2004) Some characterizations of minimal {Markov} basis for
  sampling from discrete conditional distributions. Annals of the Institute of
  Statistical Mathematics 56(1):1--17

\bibitem[{Torrie and Valleau(1974)}]{Torrie1974578}
Torrie GM, Valleau JP (1974) {Monte Carlo free energy estimates using
  non-Boltzmann sampling: Application to the sub-critical Lennard-Jones fluid}.
  Chemical Physics Letters 28(4):578 -- 581

\bibitem[{Tracy and Widom(1994)}]{tracy1994level}
Tracy CA, Widom H (1994) {Level-spacing distributions and the Airy kernel}.
  Communications in Mathematical Physics 159(1):151--174

\bibitem[{Tracy and Widom(1996)}]{tracy1996orthogonal}
Tracy CA, Widom H (1996) {On orthogonal and symplectic matrix ensembles}.
  Communications in Mathematical Physics 177(3):727--754

\bibitem[{Vogel et~al(2013)Vogel, Li, W\"ust, and
  Landau}]{PhysRevLett.110.210603}
Vogel T, Li YW, W\"ust T, Landau DP (2013) Generic, hierarchical framework for
  massively parallel {Wang--Landau} sampling. Physical Review Letters
  110:210603

\bibitem[{Vorontsov-Velyaminov et~al(1996)Vorontsov-Velyaminov, Broukhno,
  Kuznetsova, and Lyubartsev}]{vorontsov1996free}
Vorontsov-Velyaminov PN, Broukhno AV, Kuznetsova TV, Lyubartsev A (1996) Free
  energy calculations by expanded ensemble method for lattice and continuous
  polymers. The Journal of Physical Chemistry 100(4):1153--1158

\bibitem[{Vorontsov-Velyaminov et~al(2004)Vorontsov-Velyaminov, Volkov, and
  Yurchenko}]{vorontsov2004entropic}
Vorontsov-Velyaminov PN, Volkov NA, Yurchenko AA (2004) Entropic sampling of
  simple polymer models within {Wang--Landau} algorithm. Journal of Physics A:
  Mathematical and General 37(5):1573--1588

\bibitem[{Wang and Landau(2001{\natexlab{a}})}]{wang2001determining}
Wang F, Landau DP (2001{\natexlab{a}}) Determining the density of states for
  classical statistical models: A random walk algorithm to produce a flat
  histogram. Physical Review E 64(5):056101

\bibitem[{Wang and Landau(2001{\natexlab{b}})}]{wang2001efficient}
Wang F, Landau DP (2001{\natexlab{b}}) Efficient, multiple-range random walk
  algorithm to calculate the density of states. Physical Review Letters
  86(10):2050--2053

\bibitem[{Wang and Swendsen(2002)}]{wang2002transition}
Wang JS, Swendsen RH (2002) Transition matrix {Monte Carlo} method. Journal of
  Statistical Physics 106(1-2):245--285

\bibitem[{Wolfsheimer and Hartmann(2010)}]{PhysRevE.82.021902}
Wolfsheimer S, Hartmann AK (2010) Minimum-free-energy distribution of {RNA}
  secondary structures: Entropic and thermodynamic properties of rare events.
  Physical Review E 82(2):021902

\bibitem[{Wolfsheimer et al (2011)}]{Wolfsheimer2011}
Wolfsheimer S, Herms I, Rahmann S and Hartmann A K (2011).
Accurate statistics for local sequence alignment with position-dependent
scoring by rare-event sampling. BMC Bioinformatics, 12(1), 47.

\bibitem[{W{\"u}st and Landau(2012)}]{wust2012optimized}
W{\"u}st T, Landau DP (2012) Optimized {Wang--Landau} sampling of lattice
  polymers: Ground state search and folding thermodynamics of {HP} model
  proteins. The Journal of Chemical Physics 137(6):064903

\bibitem[{Yan et~al(2002)Yan, Faller, and de~Pablo}]{yan2002density}
Yan Q, Faller R, de~Pablo JJ (2002) Density-of-states {Monte Carlo} method for
  simulation of fluids. The Journal of Chemical Physics 116(20):8745--8749

\bibitem[{Yanagita and Iba(2009)}]{yanagita2009exploration}
Yanagita T, Iba Y (2009) Exploration of order in chaos using the replica
  exchange {Monte Carlo} method. Journal of Statistical Mechanics: Theory and
  Experiment 2009(02):P02043

\bibitem[{Yevick(2002)}]{Yevick2002}
Yevick D (2002) {Multicanonical communication system modeling - Application to
  PMD statistics}. IEEE Photonics Technology Letters 14(11):1512--1514

\bibitem[{Yu et~al(2011)Yu, Liang, Ciampa, and Chatterjee}]{yu2011efficient}
Yu K, Liang F, Ciampa J, Chatterjee N (2011) Efficient p-value evaluation for
  resampling-based tests. Biostatistics 12(3):582--593

\bibitem[{Zhan(2008)}]{zhan2008parallel}
Zhan L (2008) A parallel implementation of the {Wang--Landau} algorithm.
  Computer Physics Communications 179(5):339--344

\bibitem[{Zhang and Ma(2007)}]{zhang2007simulation}
Zhang C, Ma J (2007) Simulation via direct computation of partition functions.
  Physical Review E 76(3):036708

\bibitem[{Zhang and Ma(2009)}]{zhang2009counting}
Zhang C, Ma J (2009) {Counting solutions for the N-queens and Latin-square
  problems by Monte Carlo simulations}. Physical Review E 79(1):016703

\bibitem[{Zhou and Su(2008)}]{PhysRevE.78.046705}
Zhou C, Su J (2008) Optimal modification factor and convergence of the
  {Wang--Landau} algorithm. Physical Review E 78(4):046705

\bibitem[{Zhou et~al(2006)Zhou, Schulthess, Torbr{\"u}gge, and
  Landau}]{zhou2006wang}
Zhou C, Schulthess TC, Torbr{\"u}gge S, Landau DP (2006) {Wang--Landau}
  algorithm for continuous models and joint density of states. Physical Review
  Letters 96(12):120201

\end{thebibliography}

%
%
%
%
%

\end{document}